\documentclass[a4paper,twoside]{article}
\pdfoutput=1
\usepackage{pslatex}
\usepackage[a4paper]{anysize}
\marginsize{20mm}{20mm}{20mm}{20mm}
\usepackage[bf]{caption}
\usepackage{graphicx}
\graphicspath{images}
\usepackage{color}
\usepackage{epsfig}
\usepackage{amsfonts}
\usepackage{amsmath}
\usepackage{siunitx}
\DeclareSIUnit\strain{strain}
\DeclareSIUnit\UTC{UTC}
\usepackage[breaklinks]{hyperref}
\usepackage{booktabs}
\usepackage{longtable}
\usepackage{titling}
\usepackage{xpatch}
\usepackage{amssymb}
\usepackage{mathptmx}
\usepackage{bm}
\usepackage{wasysym}
\usepackage{multicol}
\usepackage{rotating}
\usepackage{fancyhdr}
\newcommand{\mytitle}{Cemented fibers as a testbed for distributed acoustic sensing (DAS)}
\newcommand{\myauthors}{Forbriger,
Münch,
Hillmann,
Rodr\'iguez Tribaldos,
Widmer-Schnidrig,
Xiao,
Rietbrock,
Strollo,
Jousset, 2026}
\usepackage{nameref}
\newcommand{\secref}[1]{'\nameref{#1}'}
\newcommand{\doiurl}[1]{\url{https://dx.doi.org/#1}}
\newcommand{\azimuth}[1]{\text{N}\SI{#1}{\degree}\text{E}}

\newcommand{\AZI}{\psi}
\newcommand{\ett}{\epsilon_{\theta\theta}}
\newcommand{\epp}{\epsilon_{\phi\phi}}
\newcommand{\etp}{\epsilon_{\theta\phi}}
\newcommand{\epsi}{\epsilon(\AZI)}
\newcommand{\eA}{\epsilon_A}
\newcommand{\eB}{\epsilon_B}
\newcommand{\eC}{\epsilon_C}
\newcommand{\Sstr}{r}

\newcommand{\Mrot}{\mathbf{M}}
\newcommand{\qmarks}[1]{`#1'}
\newcommand{\tunnellocation}[1]{\qmarks{#1}}
\newcommand{\Lanton}{\tunnellocation{Anton Gang}}
\newcommand{\Lvorstollen}{\tunnellocation{Vorstollen}}
\newcommand{\Tfiberstrain}{\qmarks{fiber strain}}
\newcommand{\Trockstrain}{\qmarks{rock strain}}
\newcommand{\Tstraintransfer}{\qmarks{strain transfer rate}}
\newcommand{\IUname}[1]{#1}
\newcommand{\IUfebus}{\IUname{A1-R}}
\newcommand{\IUquantx}{\IUname{QuantX}}
\newcommand{\IUoptodas}{\IUname{OptoDAS}}
\newcommand{\IUtreble}{\IUname{Treble+}}
\usepackage[sortcites=true,backend=bibtex,style=authoryear-comp,maxnames=2,
  giveninits=true,dashed=false,minnames=1,maxbibnames=20,
  isbn=false,natbib=true,
  uniquename=false,uniquelist=false,useprefix,alldates=long]{biblatex}
\DeclareFieldFormat[article,inbook,techreport,incollection,unpublished]{title}{#1}
\DeclareFieldFormat[article]{volume}{\textbf{#1}}
\DeclareFieldFormat[article,inbook,incollection]{pages}{#1}
\DeclareFieldFormat[article]{number}{(#1)}
\DeclareFieldFormat{doi}{%
  doi\addcolon\space
    \ifhyperref
        {\href{https://doi.org/#1}{\nolinkurl{http://doi.org/#1}}}
        {\nolinkurl{#1}}}
\xpatchbibmacro{date+extrayear}{%
  \printtext[parens]%
}{%
  \setunit*{\addcomma\space}%
  \printtext%
}{}{}

\renewbibmacro{in:}{%
\ifentrytype{incollection}{\printtext{\bibstring{in}\space}}{}}
\renewbibmacro*{volume+number+eid}{%
\printfield{volume}\space\printfield{number}\space}
\makeatletter
\newcommand{\mypbg}{%
  \begingroup
  \delimcontext{bib}%
  \edef\on@line{\on@line}%
  \@ifnextchar[%
    {\blx@printbibliography}
    {\blx@printbibliography[]}}
\makeatother
\title{\mytitle}
\author{{Thomas Forbriger}$^{1,3}$,
Felix Münch$^{1}$,
Laura Hillmann$^4$,
Ver\'onica Rodr\'iguez Tribaldos$^4$,\\
Rudolf Widmer-Schnidrig$^{2,3}$,
Han Xiao$^4$,
Andreas Rietbrock$^1$,
Angelo Strollo$^4$,
  Philippe Jousset$^4$}
\predate{\begin{flushleft}}
\postdate{\end{flushleft}}
  \date{$^1$ Geophysical Institute, Karlsruhe Institute of Technology (KIT)\\
$^2$ Institute of Geodesy, University of Stuttgart\\
$^3$ Black Forest Observatory (BFO)\\
  $^4$ GFZ Helmholtz Centre for Geosciences}
\begin{document}
\maketitle
\nocite{forbriger2025}
\section*{Abstract}
A rigid connection between the optical fiber and the rock makes
amplitudes of \Tfiberstrain\ measured with Distributed Acoustic Sensing (DAS)
equal to \Trockstrain.
We demonstrate this by running four interrogator units (IU) on a DAS testbed
with single-fiber patch cables being cemented into a groove in the concrete
floor of Black Forest Observatory (BFO).
The recorded signals are compared with the recordings of a calibrated 
Invar wire strain meter array that has been continuously in operation for the
last decades.
This way we measure \Tstraintransfer{} (ratio of \Tfiberstrain\ over
\Trockstrain) at frequencies below \SI{0.2}{\hertz}.
Waveform similarity for strong earthquake signals is high with typical values
of the normalized correlation coefficient greater than 0.95.
The \Tstraintransfer{} is close to
1 for all four IUs, while it was significantly less in a previous study with
DAS cables unreeled on the floor and loaded down by sand and sandbags, only.
At frequencies up to \SI{14}{\hertz} we make an intercomparison of IUs,
showing no significant variation with frequency.
The scatter of \Tstraintransfer{} in between channels which are spatially near
to each other in the same fiber route is about $\pm\SI{10}{\percent}$ in most
cases.
The variation of median values in between different IUs and earthquakes is
less than $\SI{5}{\percent}$.
By subtracting the common mode laser noise, which is coherent along the fiber
route, we lower the background signal level to an rms-amplitude of
$\SI{100}{\pico\strain}$ at $\SI{0.1}{\hertz}$ and $\SI{5}{\pico\strain}$ at
$\SI{1}{\hertz}$ in a bandwidth of 1/6 decade for the best cases.
This allows the detection of the marine microseisms during times of moderate
amplitude level. 

\section{Introduction}
\begin{multicols}{2}
Distributed Acoustic Sensing (DAS) is one of many distributed fiber-optic sensing
techniques by which the backscattered signal of a laser pulse sent down an optical
fiber is used to measure the conditions (deformation, temperature) along the
potentially many kilometers long fiber over intervals of a few meters or
less.
DAS, using coherent Rayleigh backscatter, as is common today, was developed in
the early 1990s.
Pioneering work was carried out by \citet{dakin1990summary}, followed by
\citet{taylor1993apparatus}.
\citet[his section 1.2.6]{hartog2017introduction} summarizes the progress made
since the early days. 
Though the term most commonly found in literature is 'DAS', \citet{hartog2017introduction} prefers the term Distributed Vibrational Sensing (DVS) as many applications
focus on rapid variations in strain, commonly referred to as dynamic strain.
For the same reason, in other contexts the term 'Distributed Dynamic Strain Sensing' (DDSS) is preferred 
\citep{jousset2025}, which better expresses the quantity discussed in the current
study.
Fields of application \citep[his chapter~9]{hartog2017introduction} include
structural health monitoring and intruder detection, but also seismic
applications.
\cite{lindsey_review_2021} and \cite{li2021literaturereview} present an
overview of fields of application in geosciences.
Many different types of optical read-out techniques have been realized in the
various types of DAS IUs \citep[his section~6]{hartog2017introduction}.
\cite{gonzales-herraez2025} 
recently pointed out that more variants tailored to specific application areas
would be conceivable but are unlikely to be developed for commercial purposes
due to the limited size of the respective instrument markets.

The specific power of DAS lies in its ability to sample deformation signals
along an optical fiber at hundreds or even thousands of locations over distances of many
kilometers with a single interrogator unit (IU).
In particular, if already deployed but unused telecommunication
infrastructures (so-called dark fibers) can be used, this is very
cost-efficient.

Although many applications rely only on the phase information in the recorded
data, the IU also provides a quantitative measure of the deformation of the
fiber, the \Tfiberstrain.
The spatial derivative of displacement, namely strain, is very sensitive to
structural heterogeneity \citep{emter1985}, which, in combination with the unprecedented spatial density of the measurements, makes DAS a powerful tool
for mapping subsurface structures such as fault zones at high resolution, as demonstrated, for example, by \citet{jousset2018}.
DAS might also provide a spatially distributed measurement of volume
strain, which leads to applications beyond the field of seismology. 
For example, with cables running in six appropriately chosen three-dimensional
directions from a single location, the full strain tensor could be composed.
From this, changes in mass density could be estimated to support
the reduction of Newtonian Noise \citep{harms2015,harms2022}, an application being explored at the planned location of the proposed Einstein Telescope \citep{et2020}.
However, the background noise in currently available DAS recordings might still be too strong for this specific application.

Such applications would hinge on the \Tfiberstrain\ being equal to the
actual rock deformation, the \Trockstrain.
\citet{forbriger2025} recently showed in a comparison with an independent
measurement of rock strain, that the ratio of \Tfiberstrain\ to \Trockstrain,
the so-called \Tstraintransfer\ is likely to be less than one. 
This presumably is due to elastic deformation within the fiber optic cable and
because of imperfect coupling of the fiber to the cable jacket and of the
cable jacket to the rock.
This was also indicated by investigations of \citet{paitz2020,lindsey2020} in
comparison with seismometer data, and \citet{reinsch2017} made a quantitative
estimation of \Tstraintransfer\ based on the elastic properties of the
different layers in the cable, based on the theoretical model by
\citet{li2006}.
Their results are summarized in more detail by \citet[their
introduction]{forbriger2025}.

\citet{forbriger2025} used several fibers in a tunnel installation of two
different types of cables, a standard telecommunication cable and a
tight-buffered cable (dedicated by the manufacturer to DAS applications).
Sections of these cables were loaded down by piled up sand and sandbags in order to improve
the coupling to the rock.
Although the waveform similarity of DAS signals compared with strainmeter
signals was very high for large amplitude signals, the fibers did not pick up
the full \Trockstrain.
The \Tstraintransfer\ turned out to vary between 0.13 and 0.53, depending
on the type of cable and installation, though no consistent effect of the
sandbags could be identified.

\citet{chien2025} calibrated the strain response of the onshore and 
offshore sections of dark fibers in a commercial telecommunication cable connecting
two islands in Puget Sound, Washington, USA.
They used a DAS interrogator (Sintela Onyx v1.0) on one of the fibers and 
an optical fiber strainmeter (OFS) on another.
Both turned out to provide highly consistent waveform data in terms of strain amplitude and phase.
Waveform similarity with respect to strain simulated from a seismometer record 
is better for the onshore data (normalized correlation coefficient, NCC 0.88)
than for the offshore data (NCC 0.48) \citep[their Fig.~7]{chien2025}
for \SIrange{0.01}{0.1}{\hertz}.
However the \Tstraintransfer\ turned out to be 0.23 only, a result which they attribute
to imperfect coupling.

The current study tests the hypothesis that the \Tstraintransfer\ being
smaller than 1 indeed is due to imperfect coupling to the rock.
For this purpose, we have cemented eight single fiber optic patch cables,
composed by only the glass fiber (core and cladding), thin coating, and tight
buffer, in a 250\,m long groove.
Four different IUs were operated simultaneously on this testbed in a kind of
huddle test in order to measure the \Tstraintransfer\ with respect to
\Trockstrain\ obtained from an array of Invar wire strainmeters and check
their signal consistency.
We measure waveform similarity, \Tstraintransfer, and the detection threshold
(background noise level) in the frequency range from 20\,mHz to 20\,Hz.

\end{multicols}
\clearpage
\section{Data}
\begin{multicols}{2}
  \raggedcolumns
\subsection{Black Forest Observatory (BFO)}
\begin{figure*}
  \includegraphics[width=\textwidth]{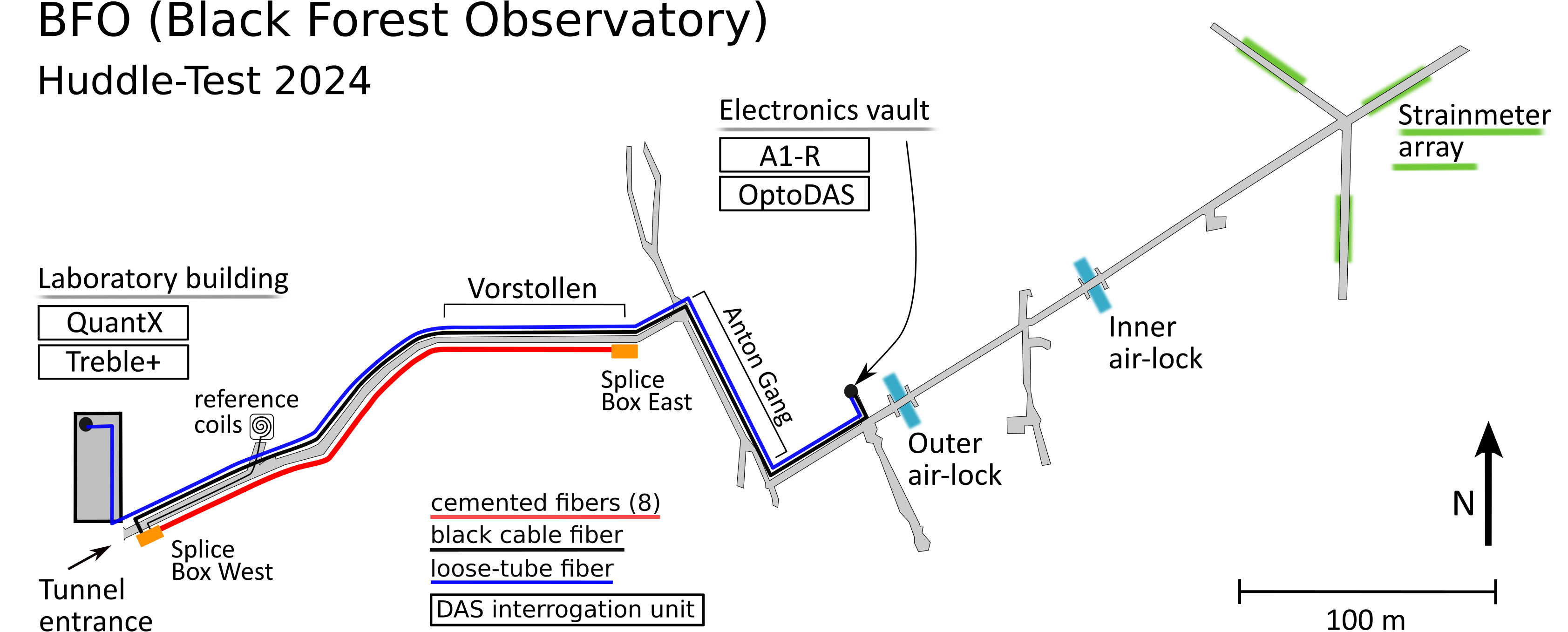}
  \caption{Floor map of the Black Forest Observatory (BFO).
  The part of the gallery used for the current experiment is shown.
  The entrance is at the western end of the tunnel.
  The overburden increases to the east and is about \SI{170}{\meter} at the
  strain meter array, which is used as a reference instrument in the current
  study.
  The eastern part hosts the majority of observatory instruments and is
  additionally protected by two air-locks.
  They are discussed by
  \citet{zuern2007a,richter1995} but have been substantially improved since
  then to act as a low-pass with a corner period of about \SI{2}{\day}.
  The DAS installation and the testbed with the cemented fibers in particular
  is located in the western part of the gallery.
  Details are shown in Fig.~\ref{fig:floor:map:test:bed}.}
  \label{fig:full:floor:map}
\end{figure*}

The Black Forest Observatory (BFO) uses a former silver mine in the
central Black Forest, Germany \citep{emter1994}.
Fig.~\ref{fig:full:floor:map} shows the part of the gallery and the
installations used in the current study.
The gallery is mostly horizontal and is excavated in granite, which
is covered with Triassic sedimentary rocks \citep{emter1994}.
Overburden increases with distance to the tunnel entrance (the western end),
is about 100\,m at the \Lanton\ and reaches 170\,m at the
\tunnellocation{Strainmeter array}.
The extent of the DAS testbed installation is marked by the
\tunnellocation{Splice Box West} and the
\tunnellocation{Splice Box East}.
We compare DAS data with recordings from the \tunnellocation{Strainmeter
array}.

\begin{figure*}
  \includegraphics[width=\textwidth]{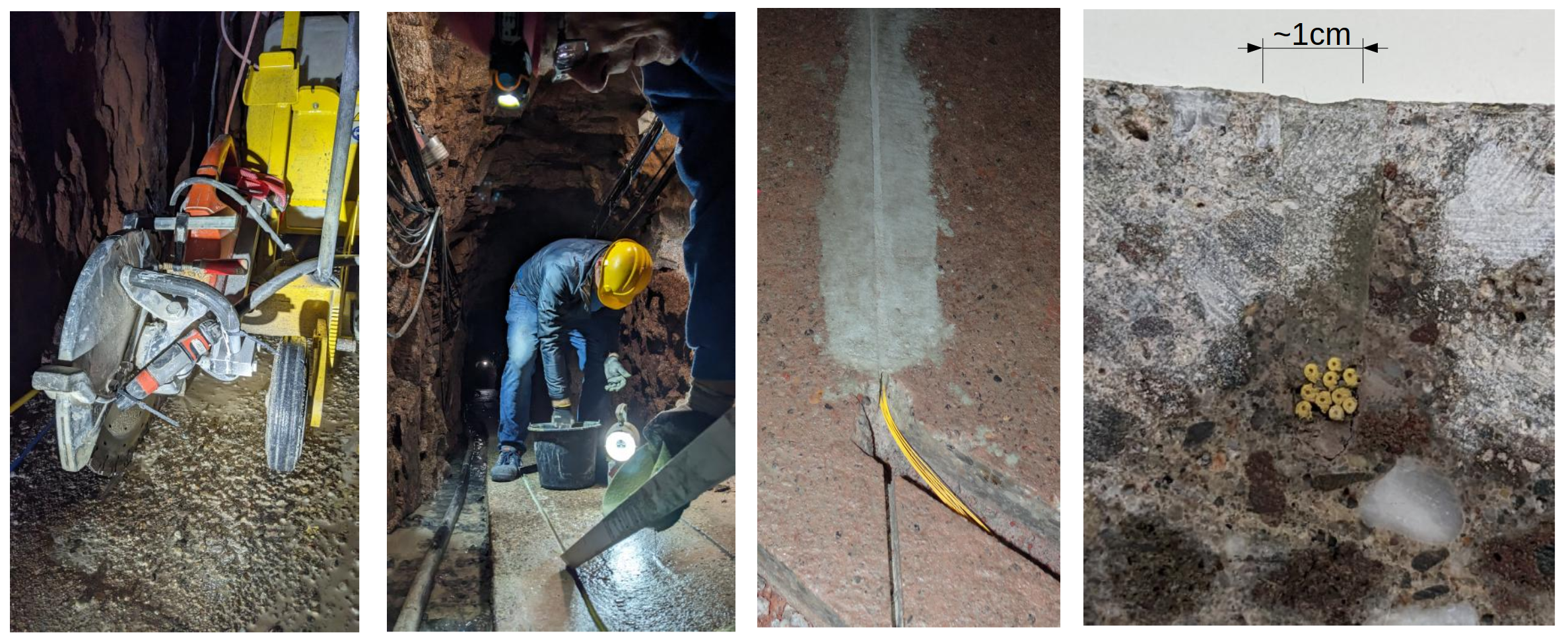}
  \caption{A \SI{10}{\milli\meter} wide and \SIrange{10}{20}{\milli\meter}
  deep groove has been cut into the concrete floor of the front part of the
  gallery (see Fig.~\ref{fig:full:floor:map}).
  Eight single-fiber patch-cables, each of them being
  \SI{0.9}{\milli\meter} thick (see Fig.~\ref{fig:fiber:layers}),
  run in parallel in the groove.
  The groove has been filled with non-shrinking cement. 
  The right-most picture shows a cross-section of a preliminary
  test-installation of 10 cables.}
  \label{fig:pictures:testbed}
\end{figure*}
\begin{figure*}
  \begin{minipage}{0.2\textwidth}
    \begin{tabular}{rl}
      $\SI{9}{\micro\metre}$ & Core \\
      $\SI{125}{\micro\metre}$ & Cladding \\
      $\SI{250}{\micro\metre}$ & Coating \\
      $\SI{900}{\micro\metre}$ & Tight buffer \\
    \end{tabular}
  \end{minipage}
  \hfill
  \begin{minipage}{0.25\textwidth}
    {\includegraphics[width=\textwidth]{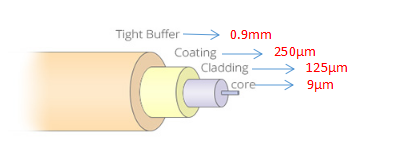}}\par
  \end{minipage}
  \hfill
  \begin{minipage}{0.5\textwidth}
    print on reel: \\
    Fiber Patch Cable, 1000m \\
    Simplex 9/125 Single Mode 0.9 mm OFNR Cable \\
    P/N: FSSM PCW20231010070038-01 \\
    www.fs.com
  \end{minipage}
  \caption{The structure of the single-fiber cable cemented into the groove.
  The cables are \SI{0.9}{\milli\meter} thick. 
  Eight of them run in parallel in the cemented groove as shown in
  Fig.~\ref{fig:pictures:testbed}.
  The tight buffer provides some protection against mechanical and chemical
  damage in the groove while still providing a tight strain coupling to the
  rock.}
  \label{fig:fiber:layers}
\end{figure*}
\begin{figure*}
  \includegraphics[width=\textwidth]{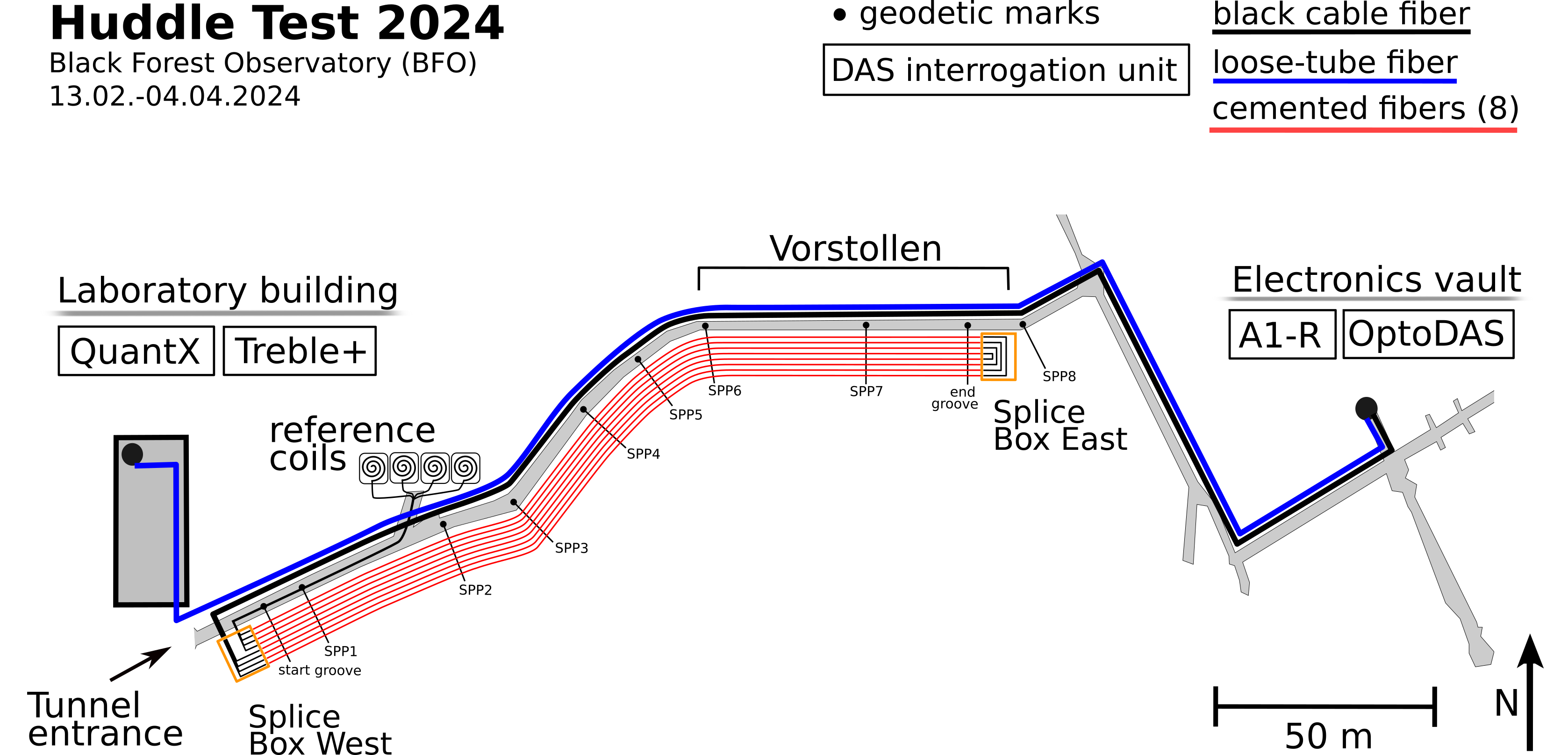}\par
  \caption{Detail of the floor map (Fig.~\ref{fig:full:floor:map}) focusing
  on the DAS testbed.
  The eight red lines schematically indicate the path of the cemented fibers
  (see Fig.~\ref{fig:pictures:testbed}). 
  SPP1 to SPP8 are marked points within the gallery, to which the channel
  offsets are aligned in tap tests.
  The reference coils are decoupled from the rock, are located in a niche and
  protected against dripping water, but are not additionally shielded.
  Due to space constraints only two IUs (\IUfebus\ and \IUoptodas) could be
  installed in the electronics vault.
  The other two units (\IUquantx\ and \IUtreble) were installed in the laboratory
  building.
  We analyze signals recorded in the straight section of the \Lvorstollen.}
  \label{fig:floor:map:test:bed}
\end{figure*}
\subsection{The testbed}
For the testbed, a \SI{10}{\milli\meter} wide and about 
\SIrange{10}{20}{\milli\meter} 
deep groove was cut into the concrete floor of the front
part of the mine (see Fig.~\ref{fig:pictures:testbed} left)
over a distance of approximately \SI{235}{\meter}.
Eight 0.9~mm thick patch-cables, consisting of core, cladding, coating, and
tight buffer only (see Fig.~\ref{fig:fiber:layers}) were placed in the groove, which was then filled with non-swelling cement.
A cross-section of a test-installation is shown in the rightmost panel of
Fig.~\ref{fig:pictures:testbed}.
The ends of these cables are collected in two wall cabinets, which we call
\tunnellocation{Splice Box West} and \tunnellocation{Splice Box East} (see
Figs.~\ref{fig:full:floor:map} and \ref{fig:floor:map:test:bed}).
The fibers are simply looped back in the \tunnellocation{Splice Box East}
such that they are combined to four pairs.
In the \tunnellocation{Splice Box West} each of the four IUs is connected to
one of these pairs, where the opposite end of the pair is connected to one of
four 'reference coils', which are placed in a niche along the gallery (see
Fig.~\ref{fig:floor:map:test:bed}).
At the very end of each fiber route the cable is wound tightly around a pen of
about \SI{8}{\milli\meter} diameter for proper termination.
Due to this small radius the light penetrates the cladding and coating and
decays away within a few centimeters such that no reflection from the end of
the fiber takes place.

As the signals from reference coils are subtracted from the channel data (see the
discussion of preprocessing below), it is essential to protect the coils from
disturbances (e.~g.\ dripping water), which otherwise would contaminate the
analyzed data.
One might even consider to provide extra thermal insulation to these coils,
which has not been done in the current experiment.
We analyze DAS data which are predominantly recorded in a straight section,
called \Lvorstollen\ (Figs.~\ref{fig:full:floor:map} and
\ref{fig:floor:map:test:bed}).
This section is about 70~m long and runs along the azimuth \azimuth{90}.

The entire fiber route beginning at the IU and ending at the end of the
reference coil appears to the IU as a long linear sensor.
Channels are typically identified by their optical distance from the IU along
the fiber route.
The actual fiber is folded several times in the gallery such that the same
point in the \Lvorstollen\ is interrogated with two channels of each fiber
route.
In order to identify these channels, we excited signals at marked locations
(SPP\,1 to SPP\,7 in Fig.~\ref{fig:floor:map:test:bed}) by tapping on the
floor (so-called \qmarks{tap-tests}) and then identifying these signals in the
recorded data.
The procedure and obtained accuracy is discussed in more detail in the
supplemental material.

\subsection{The interrogators}
\begin{table*}
  \caption{Interrogator units (IUs) and recording parameters which were chosen
  according to manufacturers recommendations.
  They were set to be as close as possible to a gauge length of
  \SI{50}{\meter}, a sampling rate of \SI{200}{\hertz}, and a channel spacing
  of \SI{10}{\meter}.}
  \label{tab:IUs}
    \begin{tabular}{ll
      S[table-space-text-post = m,table-format= 2.3,table-column-width = 2 cm]
      S[table-space-text-post = Hz,table-format= 4.0]
      S[table-space-text-post = kHz,table-format= 2.1,table-column-width = 2 cm]
      S[table-space-text-post = m,table-format= 2.2]}
    \toprule
   {interrogator,} & {recorded} & {gauge}  & {sampling} & {pulse} & {channel} 
      \\
   {manufacturer}  & {quantity} & {length} & {rate}     & {rate} & {spacing} 
      \\
    \midrule
      \IUfebus,          Febus   & strain rate & 50\,m    &  200\,Hz & 10\,kHz &  9.6\,m\\
    \IUoptodas\ C01-S, ASN       & strain rate & 40.852\,m&  250\,Hz & 50\,kHz &10.21\,m\\
    \IUquantx,        OptaSense & strain      & 51.048\,m&  200\,Hz & 10\,kHz &10.21\,m\\
    \IUtreble,       Terra 15  & velocity    & 49.822\,m& 1959\,Hz & 25.5\,kHz & 9.8\,m\\
    \bottomrule
  \end{tabular}
\end{table*}
Four different IUs, listed in Tab.~\ref{tab:IUs} were operated in parallel.
Two of them, the ASN \IUoptodas\ and the Febus \IUfebus, were installed in the
electronics vault (see Fig.~\ref{fig:floor:map:test:bed}).
Due to limited space, the other two units, the OptaSense \IUquantx\ and the Terra~15
\IUtreble, had to be installed in the laboratory building (see
Fig.~\ref{fig:floor:map:test:bed}).
The laser pulses of the latter two are transmitted through two fibers in 
a standard telecommunication cable (called 'loose-tube fiber' in
Figs.~\ref{fig:full:floor:map} and \ref{fig:floor:map:test:bed})
to the electronics vault.
Four fibers in a cable running from there
(called 'black cable fiber' in
Figs.~\ref{fig:full:floor:map} and \ref{fig:floor:map:test:bed})
transmit the pulses to the
\tunnellocation{Splice Box West} where the pulses are fed into the testbed
fibers.

The four IUs base their measurements on the analysis of changes in the phase of Rayleigh backscattered light, although they implement this in
different ways and with different reference.
Details on the interrogation principles of each IU are given in the subsections below. It is important to note that the \IUtreble\ uses an interrogation concept that differs substantially from the other IUs, with a focus on improving noise levels and sensitivity at high sampling rates. Consequently, in the current study, the \IUtreble\ is operated outside its preferred frequency range, which should be taken into account when evaluating our observations.
Despite this difference, we use these data in our evaluations as well, because they are consistent with the data from the other IUs in that they confirm the \Tstraintransfer{} near unity obtained with the cemented fibers.

Recording parameters were set to be as similar as possible for all IUs. The essential recording parameters for each unit are given in
Tab.~\ref{tab:IUs}
(more detailed parameters are listed in the supplemental material).
At a given level of phase noise, the strain noise level decreases with increasing gauge length. 
For this reason, we selected a rather large gauge length, as close to
\SI{50}{\meter} as possible for the respective IU.
With a channel interval of approximately \SI{10}{\meter} this allowed us to
use more than one channel in the \Lvorstollen.

Timing was provided to the \IUquantx\ and the \IUtreble\ through GPS antennas.
The \IUfebus\ also used a GPS signal, which was provided through an optical
link (Meinberg GOAL: GPS Optical Antenna Link) and re-radiated near the IU in
the mine.
For the \IUoptodas{} reception of the re-radiated GPS signal was unsuccessful
and time was synchronized to a 1\,PPS signal and the \SI{10}{\mega\hertz}
signal from the GOAL.
In the supplemental material, we discuss observed time deviations on the order
of milliseconds.

On 2024-03-19 the fiber routes were swapped in order to verify that there were
no obvious quality issues with the one or the other route.
The swap took place pairwise.
The route of the \IUfebus\ was swapped with that of the \IUoptodas\
and likewise between the \IUquantx\ and the \IUtreble.

\subsubsection{Febus \IUfebus}
The Febus \IUfebus\ is a $\Phi$-OTDR device providing quantitative acoustic
(vibrational) measurements through optical phase extraction. 
The system uses a single-pulse heterodyne approach for phase detection, as
described by \citet{pan2011hDVS}.
A fraction of the light produced by the laser in the IU is used as the
\qmarks{local oscillator} reference signal.
The light backscattered in the optical fiber interferes with this reference
signal, converting phase information into intensity. 
The differential phase over a gauge length provides the strain measurement,
which can also be converted to strain rate through time differentiation. 
The system allows independent and customizable gauge lengths to be acquired
simultaneously in real time.

\subsubsection{OptaSense \IUquantx}
The OptaSense \IUquantx\ interrogator unit also uses the principle of coherent
phase-sensitive optical time-domain reflectometry (phase-OTDR) with the pulsed
interrogation technique, in which short, coherent pulses of light are launched
into the fiber and coherent methods are used to record the phase changes of
the backscattered signal.
The unit offers both single-carrier and two-carrier interrogation modes, to be chosen
by the operator based on the application and desired output. The single-carrier measurement is recommended when large, high-frequency signals are expected, in a configuration in which losses along the fiber are low and high spatial resolution is important. In two-carrier mode, two independent pulses are launched to interrogate the fiber. This provides two overlapping, semi-independent backscatter patterns that result in two independent phase measurements which are combined to reduce fading. This mode is recommended for large measurement range and long gauge lengths, and handles challenging losses more effectively, but has the drawback that
there might be a small amount of crosstalk between different locations on the fibre (OptaSense, personal communication).
The current study uses the two-carrier setup, as that was the configuration available in the IU for a \SI{50}{\meter} gauge length. 
The \IUquantx\ records optical phase change between the end points of the
gauge length at a given time, which is proportional to strain over the gauge
length, and not strain-rate. 
As a result, measurements are expected to be more stable at low frequencies.
This IU uses ultrastable, very coherent lasers ensuring stability of the laser over the entire time 
the light travels through the fiber and back to the IU, minimizing laser-related drift over time (OptaSense, personal communication).

\subsubsection{ASN \IUoptodas}
The ASN \IUoptodas\ interrogator unit uses the frequency swept interrogation
technique for measuring the DAS signal reflected from the fiber. 
In this measurement principle, a linear frequency modulated (LFM) optical
signal is launched into the fiber during a relatively long period of time
(longer than for pulsed interrogation techniques), and chirped compression
techniques are used on the detected backscattered light to compress the LFM
signal into a short pulse, as described by \citet{waagard2021}. 
This approach enables launching high optical energy per interrogation period,
at limited power.
This way, the maximum energy is not limited by optical nonlinearities that 
appear if power exceeds the limit.
However, this measurement technique requires a low-noise laser.
The interrogator measures the time-differentiated optical phase between
consecutive sweeps over the gauge length, which can be converted to
longitudinal strain-rate along this section. 
Each sample then records the average strain(rate) modulation over the gauge
length.

\subsubsection{Terra15 \IUtreble}
The Terra15 \IUtreble\ interrogator is another type of pulsed interrogating
system operating on the phase-sensitive OTDR measurement principle. 
Unlike the other IUs, the \IUtreble\ measures phase difference 
between two consecutive pulses back-scattered from the same
location in the cable, and not across a \qmarks{gauge length}.
The change in phase difference over time provides the change
of fiber length up to the back-scattering point.
Its time derivative is taken as the 'fiber velocity' at the interrogated point.
For a cable perfectly coupled with the ground along its length, this
measurement of velocity should be equivalent to the vector component of ground
particle velocity along the cable axis. 
Strain-rate can be derived by post processing, either on the interrogator or, as it was done in the current study, by
the data analyst, by defining a gauge length of separation between two
interrogation points.
The difference in velocity between these points divided by the gauge length
equals strain-rate.

Another significant difference between the Terra15 \IUtreble\ and the other interrogators is the laser used to launch light into the fiber. 
The \IUtreble\ is purposely designed to obtain high sensitivity and low noise
levels at high frequencies ($>\SI{50}{\hertz}$). 
To achieve that, a broadband laser is used, which results in more accuracy at
high frequencies, less signal fading at certain frequencies, and improved
location precision, but is prone to higher noise levels at lower frequencies (Terra15, personal communication).  
As stated above, in the current study the \IUtreble\ was operated at frequencies outside its domain of
optimal performance, which shall be kept in mind.
\subsection{Vibration Isolating Platform}
In order to decouple the IUs from floor vibrations as much as possible,
the \IUfebus, \IUquantx, and \IUtreble\ were placed on \qmarks{Minus K
Technology 100BM-8 Vibration Isolating Platforms}.
These platforms are completely passive.
They use negative-stiffness mechanisms
and an arrangement of springs in order to isolate six degrees of freedom.

The horizontal natural frequency of these tables is between \SI{1.5}{\hertz} and
\SI{2.5}{\hertz} (depending on the payload). The vertical natural frequency is
\SI{0.5}{\hertz} (independent of the weight).
At higher frequencies, the platform does not follow the ground motion due to
the inertia of its payload.

The \IUoptodas{} was shipped in a rugged
field rack too heavy for the vibration isolation table, and the interrogator
was not dismantled to be placed on such a table. 
Instead, the \IUoptodas{} was kept in its rugged case throughout the tests. 
For the \IUfebus\, the weight had to be increased by placing brass cylinders on
top of the casing, to provide sufficient inertia.
Despite this arrangement, the platform could not be properly adjusted and tended to drift against its stops.
For both the \IUoptodas\ and the \IUfebus, the vibration isolation was less
essential than for the other two IUs, as the floor in the mine provides a
more stable and less vibrating support than the wooden floor of the
laboratory.

\subsection{The strainmeter array}
We compare the DAS recorded strain with data from the BFO strainmeter array,
which is understood to represent \Trockstrain.
The array (see Fig.~\ref{fig:full:floor:map}) consists of three 10\,m long,
horizontal, Invar-wire strainmeters, which are well calibrated for the purpose
of tidal research.
The SEED codes for the instruments are
II.BFO.00.BSA,
II.BFO.00.BSB, and
II.BFO.00.BSC.
Data is available through the \cite[][\doiurl{10.7914/SN/II}]{sio1986} via the
EarthScope Data Center
(IRISDMC)\footnote{\url{https://www.earthscope.org/data/}}.
The design of these instruments is based on the instruments by
\cite{king1976} and are discussed in more detail by \cite{zuern2015}.
Depending on the method used, the calibration accuracy can be better than
\SI{2}{\percent} or \SI{5}{\percent} for calibration with respect to
'Crapaudines' or synthetic tides, respectively.
Details of the calibration mechanism are discussed by
\citet[their section S2.4]{forbriger2025}.

Signals recorded by these instruments are used only up to frequencies below
\SI{1}{\hertz}.
At higher frequency they show a linear parasitic sensitivity to vertical
ground acceleration because of the inertia of the pick-up system.

Other studies use seismometer data to compare with DAS signals.
By using \Trockstrain\ directly measured with strainmeters we can avoid this.
The use of seismometer data hinges on a plane-wave assumption, which in
heterogeneous settings and with crustal wave propagation is likely to be
incorrect.
\citet[their sections 'Comparison with seismometer data' and
S4]{forbriger2025} discuss this and point out further complications due to
local strain-strain coupling in their supplementary section~S2.5.

\subsection{Analyzed time windows}
The recording of the experiment covers the period from 2024-03-05 to
2024-04-04.
During this one month, we find three earthquakes which provide 
a signal-to-noise ratio of about ten for at least three of the IUs.
These signals cover a frequency range of about 20\,mHz to 20\,Hz
and are listed in Tab.~\ref{tab:earthquakes}.
Due to their different magnitude and epicentral distance, each of the
earthquakes covers a different frequency band in the following analyses.

Additionally, we use time windows free of earthquakes to estimate the
background signal level.
Here we focus on a time window with moderate to low marine
microseism amplitude.

\begin{table*}
  \caption{Earthquake signals used for the analysis.
  We estimate the signal-to-noise ratio (SNR) by comparing amplitude spectra
  of the earthquake signal recorded with the \IUfebus\
  to spectra of a sequence of the same length preceding the earthquake.
  BAZ: backazimuth}
  \label{tab:earthquakes}
  \begin{tabular}{llr
    S[table-space-text-pre = M,table-format= 3.1]
    S[table-space-text-post = km,table-format= 2.1]
    S[table-space-text-post = Hz,table-format= 1.2,table-column-width = 12 mm]
    c
    S[table-space-text-post = Hz,table-format= 2.1,table-column-width = 12 mm]}
    \toprule 
    {earthquake origin time} &  {location} & {BAZ} & {magnitude} &  
    {distance} &  
    \multicolumn{3}{c}{bandwidth (SNR $\apprge 10$)}\\ 
    \midrule  
    2024-04-02 23:58:09 UTC &  Hualien City, Taiwan & \azimuth{57} & M\,7.4
    & 86.6\,$\si{\degree}$ 
    & 0.02\,Hz & {--} & 0.2\,Hz \\ 
    2024-03-27 21:19:37 UTC & Tolmezzo, Italy & \azimuth{121} & M\,4.5
    & 3.7\,$\si{\degree}$ 
    & 0.2\,Hz & {--} & 8.0\,Hz\\ 
    2024-03-22 05:31:50 UTC &  Albstadt, Germany & \azimuth{105} & M\,2.8
    & 59\,km 
    & 3.0\,Hz  & {--} & 18.0\,Hz\\ 
    \bottomrule
  \end{tabular}
\end{table*}

\end{multicols}
\section{Analyses}
\begin{multicols}{2}
We would like to begin with a statement of our intentions.
We use four IUs in order to corroborate the properties of the testbed and the
\Tstraintransfer{} in particular, by validation with IUs which implement
different techniques.
We do not intend to rank the IUs in any way.
The recording parameters and analyzed frequency bands for the four IUs have
been chosen as similar as possible, which in turn means that these parameters
might not be optimal for each of the IUs.
In most of the data we find details, which we still do not fully understand.
Some of them might be due to the recording conditions and be caused by
water occasionally dripping from the mine ceiling onto the fibers, 
by temperature fluctuations, or by the different
installation conditions for the IUs themselves.
These details need not represent a property of the respective IU.
Where disturbances are obvious, we have not used the affected time windows.
In the following, we discuss the gross features of the DAS data, in particular
with respect to the coupling of the fiber to the rock and the potential of this technology to record rock strain.
We will mention similarities and differences between data from different IUs,
but like to remind the reader that these might be specific to the
particular setup.
Only selected data examples are shown, because we do not intend a 1:1
comparison.

\subsection{Pre-processing of DAS data}
Raw data recorded by the different IUs differ in size per file and duration of
file from \SI{1.1}{\mega\byte} for \SI{10}{\second} of \IUoptodas\ data to 
\SI{2}{\giga\byte} for \SI{50}{\minute} of \IUtreble\ data.
In the first step, we select the time period to be analyzed from the raw data,
concatenate the files and downsample 
(an 8th order Chebyshev type I anti-alias filter is applied)
to \SI{20}{\hertz} or \SI{200}{\hertz} depending on the frequency band of
interest.
The polarity is consistently set to make rock extension positive.
For the \IUfebus{} and \IUtreble{},
where the IUs use other values than $n=1.47$ and $\xi=0.78$ for the refractive
index and the opto elastic factor for fused silica, respectively, we adjust the scaling
accordingly to make data consistent (see the respective discussion in the
supplement).
Samples are scaled to 
strain rate in \si{\nano\strain\per\second},
or in the case of the \IUquantx\ to strain in \si{\nano\strain}, and then are
written to intermediate files in a common format.
Upon reading these data for the subsequent analysis, time series in strain
rate are converted to strain by integration (cumulative sum in the time
domain multiplied by sampling interval).
Prior to the integration we cast the data to double precision, where IUs
do not provide them as 64bit floats.
This is necessary to avoid significant loss of numerical precision in cases of
strong strain drift.

Common mode laser noise \citep[][their section 2.6 Optical
Noise]{lindsey2020}, which is coherently present on all channels, is a typical
issue of DAS systems.
In our experiment, noise of this character typically dominates at lower frequencies in the
investigated band.
In order to remove this noise, we employ the reference coils mentioned above. These reference coils (see Fig.~\ref{fig:floor:map:test:bed}) are decoupled from
the rock.
Thus, signals extracted from channels on these coils are expected to contain
only system-generated noise.
By subtracting the average time series of the channels on the reference coil
from the signal obtained from a channel in the groove, coherent laser noise
can be reduced.
This can improve the signal-to-noise ratio by up to \SI{20}{\deci\bel} at
\SI{0.1}{\hertz}.
At the same time this makes the reference coils a critical component.
Any additional noise (dripping water, temperature fluctuations, etc) in the reference channels potentially deteriorates the signals under investigation.
\citet{diazmesa2023} discuss stacking of signals recorded in co-located fibers
in order to improve the signal to noise ratio.
We have not applied this in order to be able to analyze signals from co-located fibers
separately.

An example of improved signal-to-noise ratio is shown in Fig.~\ref{fig:spec:QuantX:20240326} and is discussed in
section \secref{sec:background:signal}.
Only for the \IUtreble, which is operated outside its preferred frequency
range, incoherent noise dominates all channels such that this
measure does not improve the signal quality and we do not apply this procedure
to \IUtreble{} data.
Time domain examples of the procedure are discussed by
\citet[their section S2.3 and figures S26 and S27]{forbriger2025} and are
shown in the supplement of the current contribution.

Finite phase shifts between waveforms become discernible
for frequencies of about \SI{10}{\hertz}, which are present in the data from
the Albstadt earthquake (Tab.~\ref{tab:earthquakes}).
All IUs are synchronized to UTC time by GNSS reception.
Nonetheless, we observe consistent time shifts of
\SIrange{6}{40}{\milli\second} between different units.
Additionally, we see the faint signature of propagation delay, which can be
\SI{2.5}{\milli\second} over the channel distance of \SI{10}{\meter}
for a surface wave at \SI{2.5}{\kilo\meter\per\second}
wave speed and incidence from the East (Albstadt) along the \Lvorstollen\
(Fig.~\ref{fig:floor:map:test:bed}).
Phase shifts would systematically reduce the measures we use ('normalized
correlation coefficient' and 'regression coefficient', see below).
For this reason we determine the time shifts by cross-correlating the
waveforms and then correct for the time shift, by accordingly resampling the
waveform with respect to the reference signal in the respective analysis (at
least for the signals of the Albstadt earthquake).

For the purpose of waveform comparison, we apply a bandpass by separate
Butterworth high- and low-pass filters. 
Both are 4th order filters. 
Below, we will specify only the filter frequencies for the respective case.
For regression and correlation analysis all involved time series are resampled 
to the highest sampling rate involved by Lanczos interpolation 
as implemented in ObsPy \citep{krischer2015}.

\subsection{Linear strain from the strainmeter array as a reference}
We compare DAS recordings with recordings of the three strainmeters
(eastern end of the tunnel in Fig.~\ref{fig:full:floor:map}).
Linear strain in the azimuth $\AZI=\azimuth{90}$ of the \Lvorstollen\ 
(Figs.~\ref{fig:full:floor:map} and \ref{fig:floor:map:test:bed})
is obtained by a linear combination of the three strain values 
$\eA$, $\eB$, and $\eC$ from the three array-instruments
with azimuths of \azimuth{2}, \azimuth{60}, and \azimuth{300}, respectively. 
They correspond to the SEED channel names BSA, BSB, and BSC.

From the components $\ett$, $\epp$, and $\etp$ of the strain tensor in 
the horizontal plane,
linear horizontal strain in an arbitrary azimuth $\AZI$ is
\begin{equation}
  \epsi=\ett\cos^2(\AZI)+\epp\sin^2(\AZI)-\etp\sin(2\AZI)
\end{equation}
as given by \citet[][their eq.~2]{zuern2015}.
Based on the strainmeter recordings this is
\begin{equation}
  \epsi=
    \begin{pmatrix}
  \cos^2(\AZI) \\ \sin^2(\AZI) \\ -\sin(2\AZI)
  \end{pmatrix}
  \,
  \Mrot^{-1}\,
  \begin{pmatrix}
    \eA \\ \eB \\ \eC
  \end{pmatrix},
  \label{eq:linear:strain}
\end{equation}
where the rotation matrix
\begin{equation}
  \Mrot^{-1}
  = 
    \begin{pmatrix}
     1.002  & -0.041  &  0.040      \\
    -0.334  &  0.680  &  0.653      \\
     0.000  & -0.577  &  0.577
  \end{pmatrix}
\end{equation}
is the inverse of
\begin{equation}
  \Mrot=
  \begin{pmatrix}
    \cos^2(2^\circ) & \sin^2(2^\circ) & -\sin(2\times 2^\circ) \\
    \cos^2(60^\circ) & \sin^2(60^\circ) & -\sin(2\times 60^\circ) \\
    \cos^2(300^\circ) & \sin^2(300^\circ) & -\sin(2\times 300^\circ) \\
  \end{pmatrix}.
\end{equation}

The strainmeters are at a distance of about \SI{350}{\meter}
from the \Lvorstollen{} (Fig.~\ref{fig:full:floor:map}).
At frequencies below \SI{0.2}{\hertz}, where the strainmeter data is used,
wavelength is much larger than this distance even for the surface waves
which propagate at about \SI{3}{\kilo\meter\per\second}.
For this reason we expect the strain waveform
at both locations to be practically the same.
The strain amplitude, however, is altered due to the local surface topography
at BFO (strain-strain coupling resulting from stress-free surfaces).
\citet{zuern2015} report an amplitude reduction to \SI{58}{\percent} for the
\azimuth{60} strainmeter, which appears more or less consistent with the
reduction to \SI{67}{\percent} in \azimuth{90} as computed by 2D finite
element analysis as reported by \citet{emter1985}.
While this effect generally applies to the location of the strainmeter array
as well as the \Lvorstollen{}, we expect the \azimuth{90} strain amplitudes to
be slightly (\SIrange{5}{10}{\percent}) larger at the \Lvorstollen{}, based on
the results by \citet[their figure~5]{emter1985}.
Further details are discussed by \citet[their section S2.5]{forbriger2025}.

\subsection{Background signal}
\label{sec:background:signal}
\begin{figure*}
    {\includegraphics[clip,trim=0 0 0 34,height=66mm]{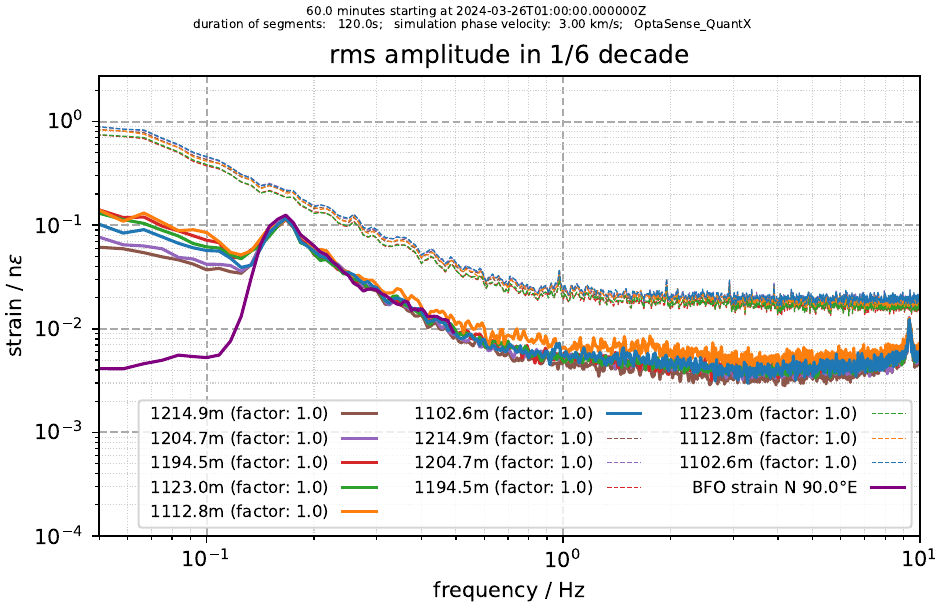}}
  \hfill
    {\includegraphics[clip,trim=0 0 0 34,height=66mm]{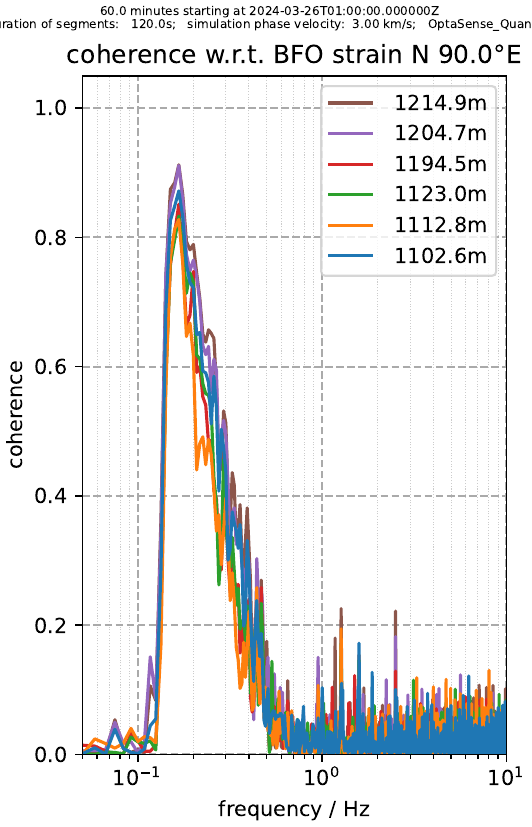}}
    \caption{Analysis of the background signal on 2024-03-26
    \SIrange{1}{2}{\UTC} for recordings of the \IUquantx:
    rms amplitude in 1/6 decade \citep[left, see section
    4.1 by][]{forbrigerPSILN2023} and magnitude-squared coherence 
    \citep[their eq.~2]{carter1973} with respect to the BFO strainmeter data
    (right).
    The corners of the applied bandpass are at \SI{20}{\milli\hertz} and
    \SI{15}{\hertz} and thus lie outside the displayed range.
    We compare signals from six channels (indicated by their offsets) to the
    \Trockstrain\ in \azimuth{90} as obtained from the BFO strainmeter array
    (purple curve).
    The curve for the latter is truncated at \SI{0.5}{\hertz}, because
    of the parasitic sensitivity to vertical ground motion at higher
    frequencies.
    The dashed lines in the left diagram show the signal level of the raw DAS
    data.
    The solid lines are the levels after subtracting the average of the signal
    on the reference coil in the time domain.
    Signal amplitudes are not further scaled to correct for \Tstraintransfer\
    (factor: 1.0).
    They follow the curvature of the purple line (\Trockstrain) at 
    \SIrange{0.14}{0.5}{\hertz}.
    The signal amplitude of the marine microseisms is at a rather low level of
    \SI{0.1}{\nano\strain} in this example.
    Nevertheless the signal is detected by the DAS, which is also confirmed by
    the coherence being raised to 0.8 in the microseism frequency band.}
    \label{fig:spec:QuantX:20240326}
\end{figure*}
Fig.~\ref{fig:spec:QuantX:20240326} shows the analysis of the background
signal on a day with low marine microseism amplitude.
The dashed lines show the raw signal levels of six channels in the
\Lvorstollen\ for the \IUquantx.
By subtracting the average of the signals on the reference coil in the time
domain, the signal level is lowered to the solid lines, which demonstrates the
effectiveness of this procedure to remove coherent noise.
This measure is similarly effective for the \IUoptodas\ and \IUfebus\ data,
though they arrive at slightly different signal levels (see additional
diagrams in the supplement).
These three IUs detect the marine microseism signal at about
\SIrange{0.14}{0.3}{\hertz}, which is also indicated by the increased level of
coherence (Fig.~\ref{fig:spec:QuantX:20240326} right panel).
The signal level in Fig.~\ref{fig:spec:QuantX:20240326} (left) is consistent
with the level of \Trockstrain\ measured with the BFO strainmeter array for
\SIrange{0.14}{0.5}{\hertz}.
The raw \IUtreble\ data is at the level of the raw data of the other three,
but is dominated by incoherent noise.
The subtraction of the reference coil signal hence does not lower the
background level.
We should remind ourselves that the \IUtreble\ is operated outside its optimal
use case, which would be at frequencies larger than investigated here.

We have run this type of analysis as shown in Fig.~\ref{fig:spec:QuantX:20240326}
for all IUs at times of moderate signal level of the marine microseisms
before and after the swapping of fiber routes on 2024-03-19.
Analysis results for the 2024-03-08 and the 2024-03-26 are shown in the
supplemental material.
By this we have convinced ourselves that the results regarding background
level are largely independent of the fiber route used. 

\subsection{Waveform similarity}
\begin{figure*}
  \includegraphics[trim=0 0 0 60,clip,width=\textwidth]{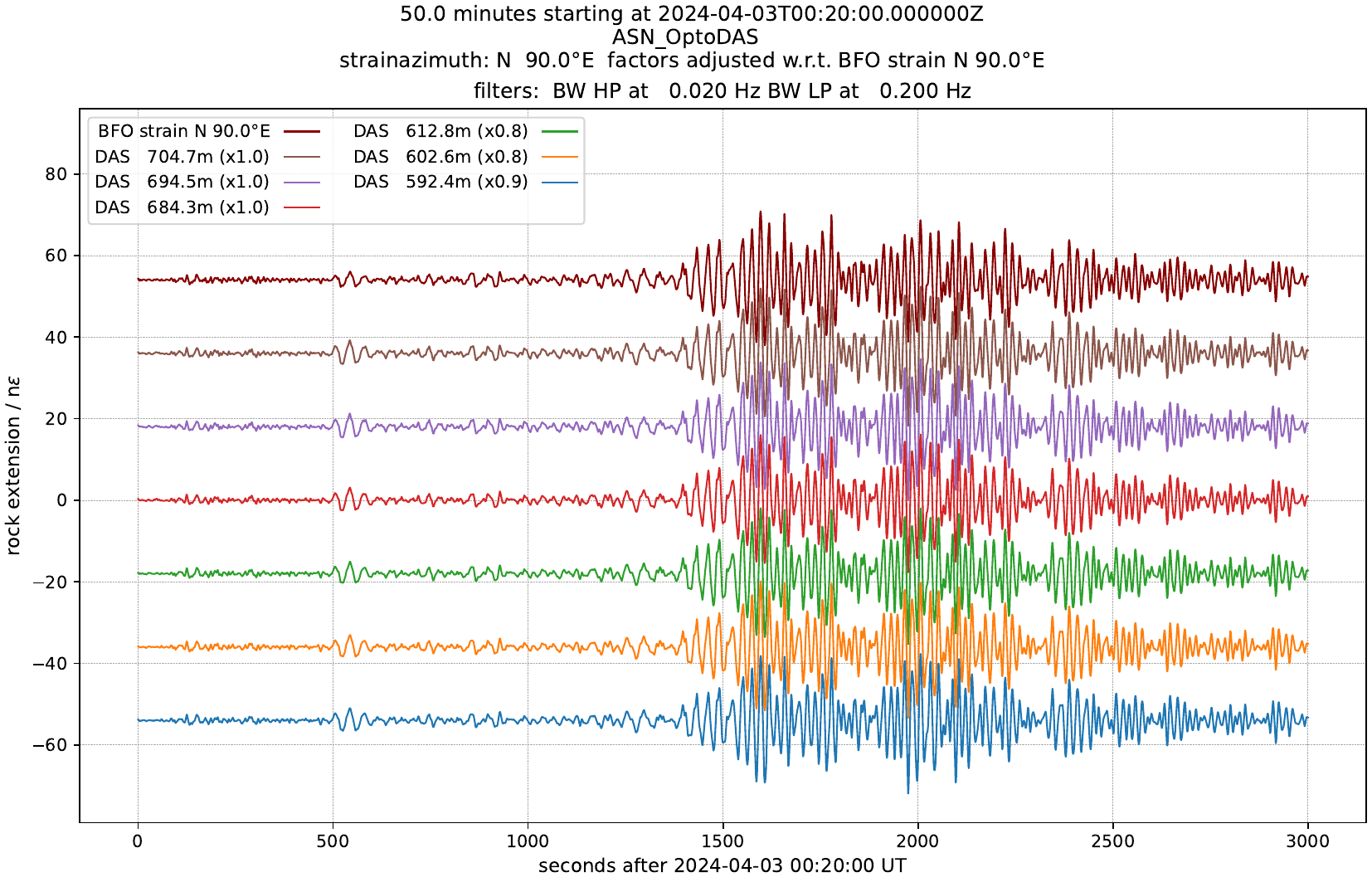}
  \caption{Waveform comparison for the Taiwan earthquake (see
  Tab.~\ref{tab:earthquakes}) recorded with the \IUoptodas{}.
  The traces are vertically shifted for better visibility of details.
  The S-phase arrives at about 2024-04-03 00:21 UTC.
  The P-phase (not shown) does not exceed the noise level.
  Rayleigh waves arrive after \SI{1300}{\second} on the time scale.
  All signals are band-passed to frequencies from \SI{20}{\milli\hertz} to
  \SI{200}{\milli\hertz}.
  The top trace is the signal derived from the BFO strainmeters in \azimuth{90}
  azimuth.
  The other six traces are recorded by the \IUoptodas{} in the \Lvorstollen{}.
  Optical channel offset and amplitude scaling factors are given in the
  legend, where the scaling factors are the reciprocal \Tstraintransfer\ (see
  Tab.~\ref{tab:OptoDAS:Taiwan:coefficients}).}
  \label{fig:man:Taiwan:OptoDAS}
\end{figure*}

For transient signals, we measure the waveform similarity by the 
normalized correlation coefficient (NCC)
\begin{equation}
  c=\frac{\sum\limits_k \, x_k\,y_k}{\sqrt{\sum\limits_k
  x^2_k}\,\sqrt{\sum\limits_k y^2_k}},
  \label{eq:normalized:correlation}
\end{equation}
where $x_k$ and $y_k$ are the samples of the one and the other time series,
respectively.
A waveform example with signals recorded after the Mw\,7.4 Taiwan earthquake
(see Tab.~\ref{tab:earthquakes}) is shown in Fig.~\ref{fig:man:Taiwan:OptoDAS}.
Waveform differences are hardly noticeable as the NCC for the displayed traces
is 0.98 and larger (see Tab.~\ref{tab:OptoDAS:Taiwan:coefficients}).
\begin{table*}
  \caption{Normalized correlation coefficient (NCC) and \Tstraintransfer\ for
  the data example in Fig.~\ref{fig:man:Taiwan:OptoDAS}.}
  \label{tab:OptoDAS:Taiwan:coefficients}
  \begin{tabular}{lrrrrrr}
    \toprule
    channel
    & \SI{592.4}{\meter}
    & \SI{602.6}{\meter}
    & \SI{612.8}{\meter}
    & \SI{684.3}{\meter}
    & \SI{694.5}{\meter}
    & \SI{704.7}{\meter}\\
    \midrule
    NCC & 0.98 & 0.99 & 0.99 & 0.98 & 0.98 & 0.98\\
    \Tstraintransfer & 1.17 & 1.21 & 1.19 & 0.96 & 1.02 & 1.03\\
    \bottomrule
  \end{tabular}
\end{table*}

\begin{figure*}
  \begin{minipage}{0.3\textwidth}
    {\includegraphics[width=\textwidth]{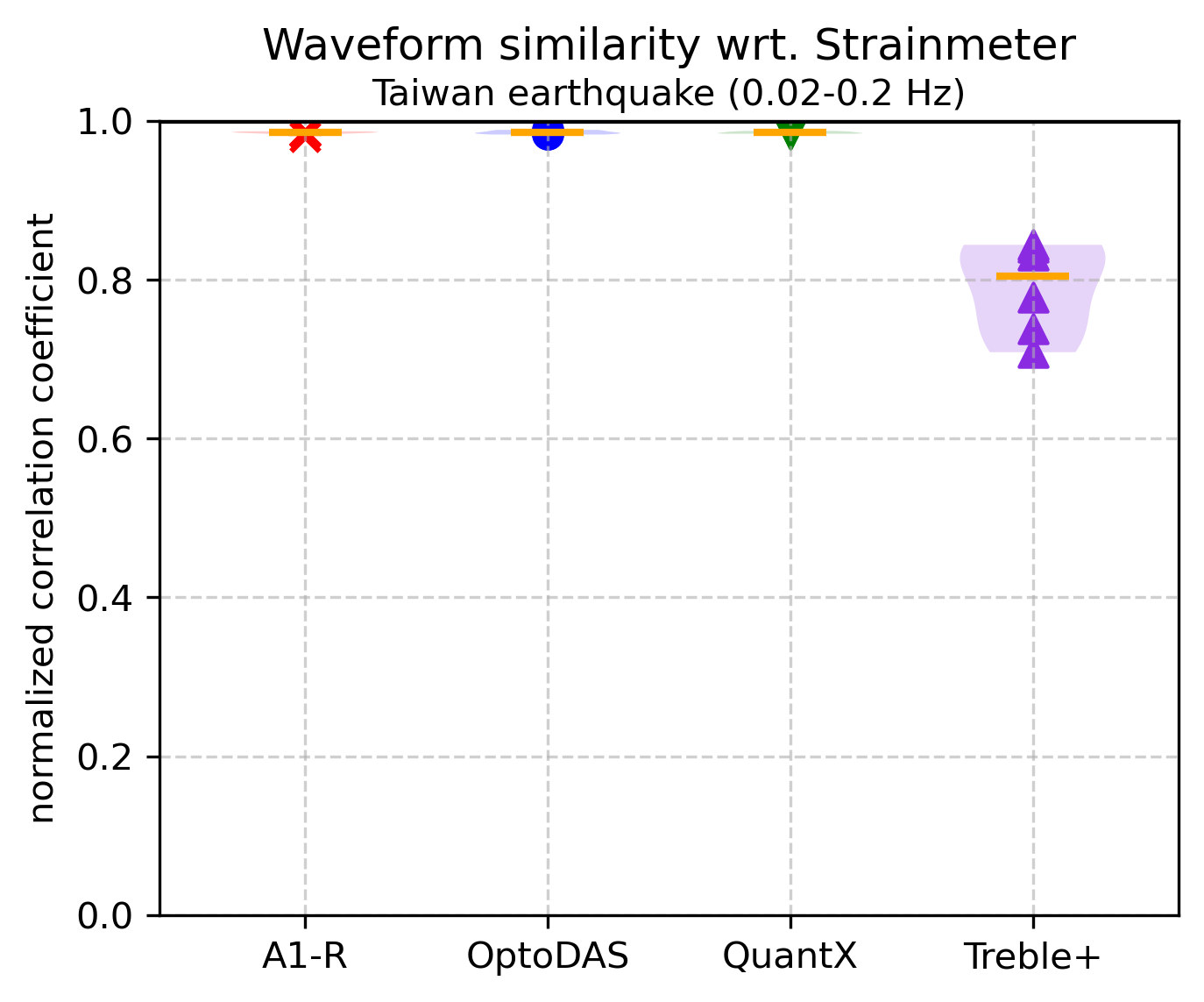}}
  \end{minipage}
  \hfill
  \begin{minipage}{0.65\textwidth}
    {\includegraphics[width=\textwidth]{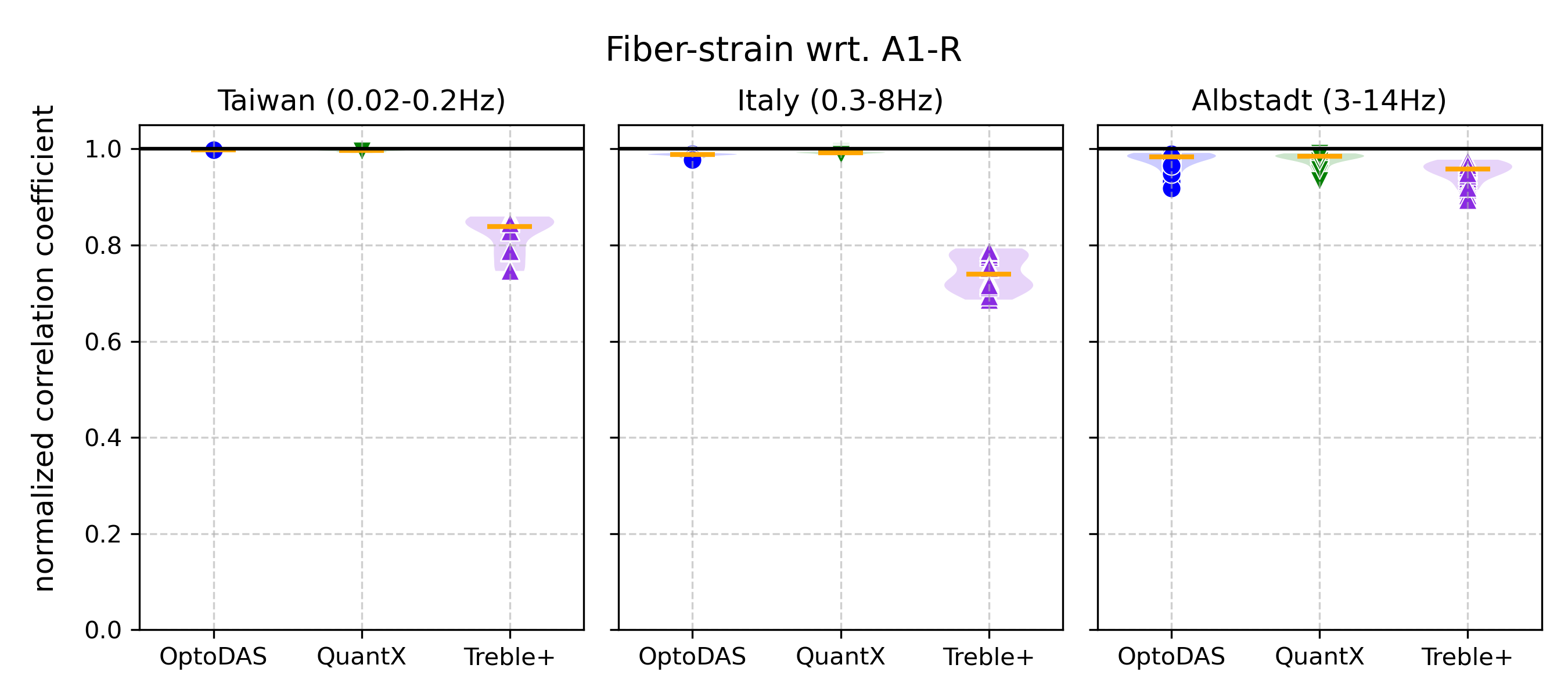}}
  \end{minipage}
  \par
  \caption{Compilation of values for the normalized correlation coefficient
  (NNC) with respect to the BFO strainmeters (left) and with respect to the
  \IUfebus\ (right).
  For the frequency band of \SIrange{0.02}{0.2}{\hertz} the DAS data can
  be compared with the BFO strainmeter data.
  For each of the IUs six channels along the \Lvorstollen\ are selected.
  The intercomparison of DAS signals (right) is also meaningful for
  frequencies larger than \SI{0.2}{\hertz}.
  To account for variations in the \IUfebus{} signals between channels, we
  compute the NCC with respect to each of the six \IUfebus{} channels,
  which results in 36 data points in the diagrams on the right (six \IUfebus{}
  channels times six channels for the IU under investigation).
  The scatter is so small that most of these 36 points are not
  distinguishable.
  The durations of the analyzed waveforms are \SI{2820}{\second},
  \SI{150}{\second}, and \SI{25}{\second} for the 
  Taiwan, Italy, and Albstadt earthquakes (Tab.~\ref{tab:earthquakes}),
  respectively.
  In each case the window covers the P-wave arrival as well as the surface
  wave train.
  The orange horizontal bars mark the median values.
  The kernel density estimates that define the edges of the violin plot
  illustrate the spread in the distribution of values.}
  \label{fig:NCC:compilation}
\end{figure*}

We have run the same analysis for the other IUs.
The values for the NCC with respect to the strainmeter signal are displayed in
Fig.~\ref{fig:NCC:compilation} (left).
They are 0.98 and larger, except for the \IUtreble, for which they still are
0.7 and larger.
The \IUtreble\ signals do not benefit from the reference coils and thus their
  signal-to-noise ratio (SNR) is lower.
For the \IUfebus, \IUquantx, and \IUoptodas\ the waveforms in the investigated
time window and frequency band are practically identical to the
strainmeter signals.

At higher frequencies a comparison with the strainmeter signal cannot be done
reasonably, because the strainmeter shows a parasitic sensitivity to vertical
  ground motion above \SI{1}{\hertz}.
  Thus, for the analysis of the waveforms recorded after the Italy and Albstadt earthquakes (see
  Tab.~\ref{tab:earthquakes}), we choose the \IUfebus\ as a reference.
The values for the NCC are displayed in Fig.~\ref{fig:NCC:compilation} (right)
and most are close to 1.
The effect of the lower SNR for the \IUtreble\ is obvious.
The SNR for the Albstadt and the Italy earthquake is generally smaller, which
is indicated by a slightly stronger scatter and reduced values of the NCC, in
particular for the Albstadt earthquake, though the median values for the
\IUquantx{} and the \IUoptodas{} are still close to 1.
The \IUtreble\ shows a higher NCC for the Albstadt earthquake than for the
other two earthquakes.
This might indicate the better waveform quality, which could be obtained in
its preferred domain of operation at even higher frequency.

\subsection{Strain transfer rate}
\begin{figure*}
  \begin{minipage}{0.3\textwidth}
    {\includegraphics[width=\textwidth]{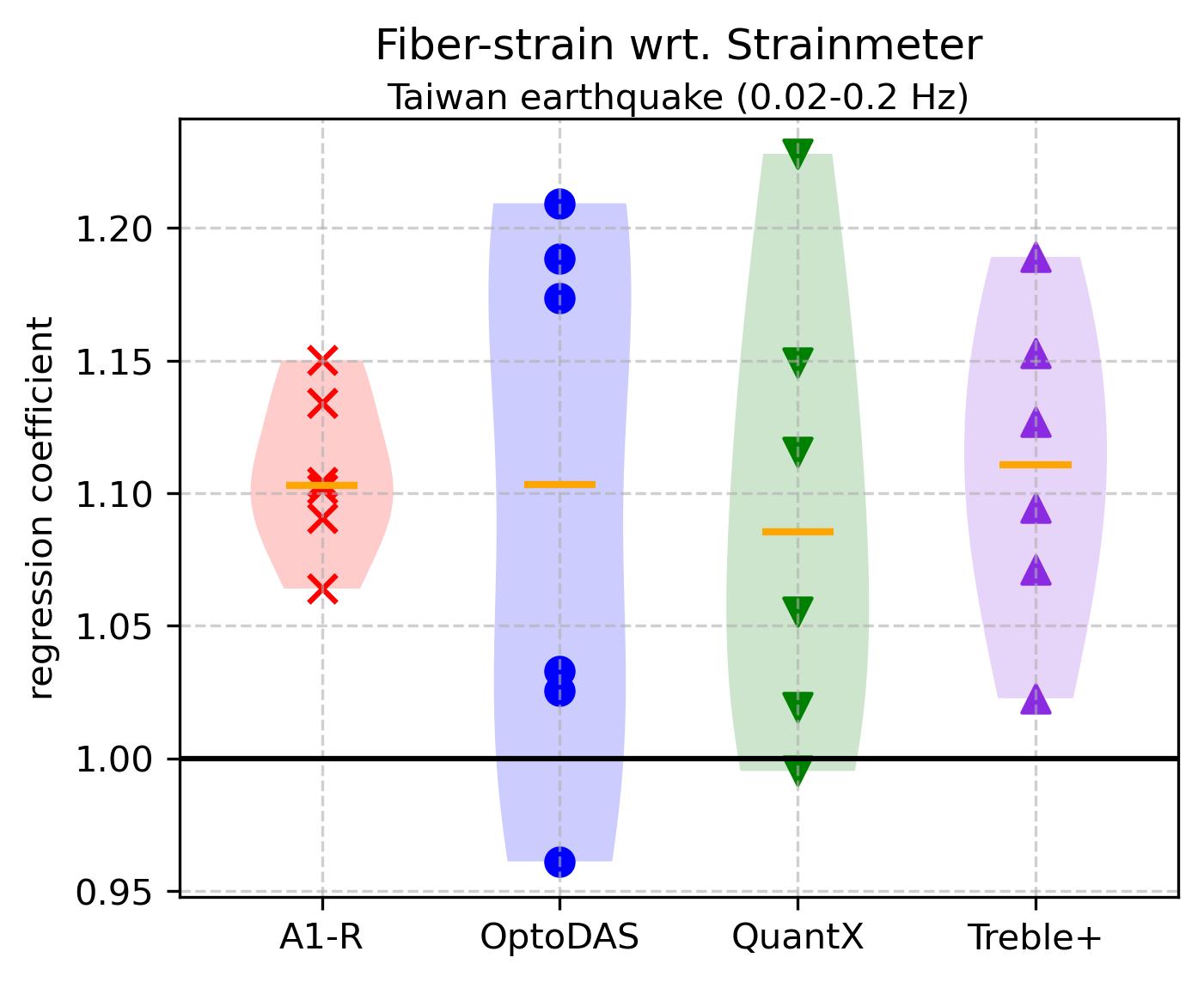}}
  \end{minipage}
  \hfill
  \begin{minipage}{0.65\textwidth}
    {\includegraphics[width=\textwidth]{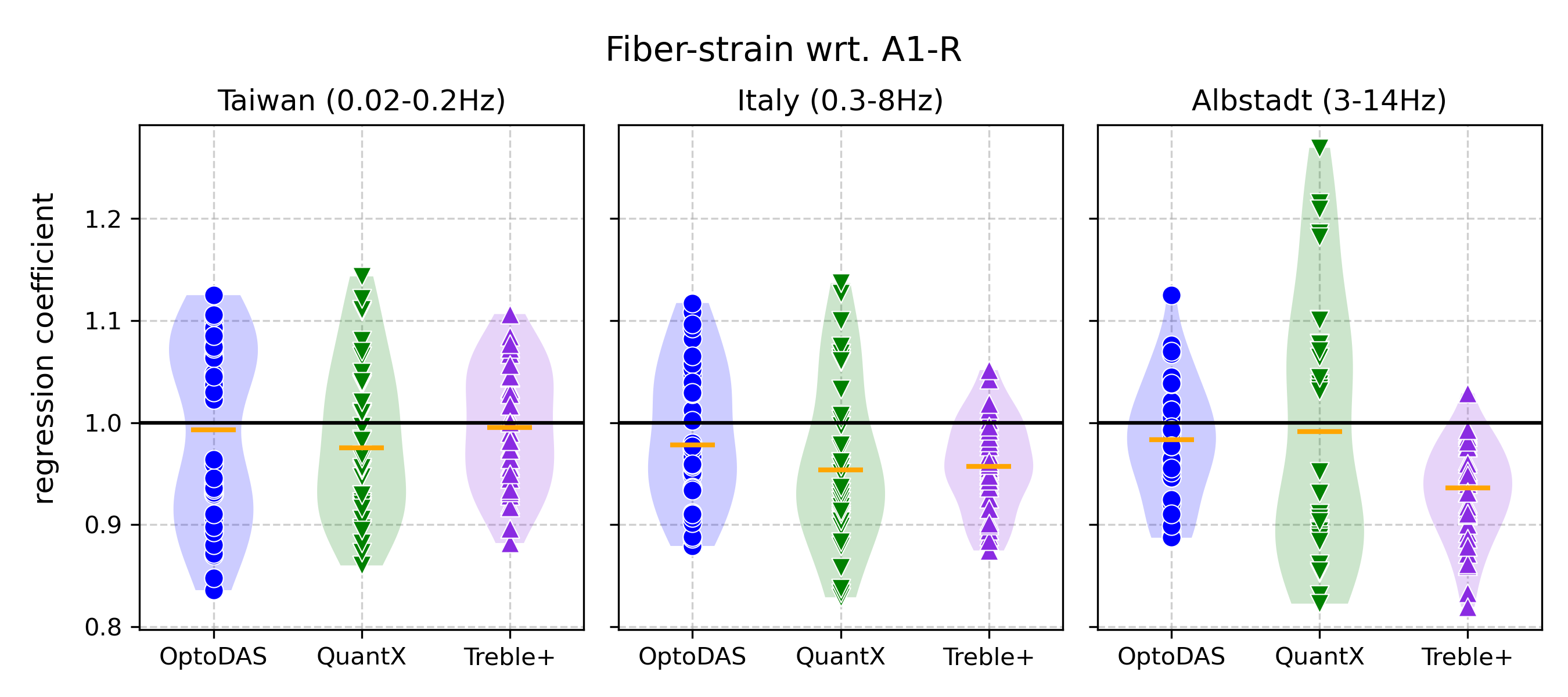}}
  \end{minipage}
  \par
  \caption{Compilation of values for
  \Tstraintransfer\ (left) and regression coefficient
  with respect to the \IUfebus\ (right).
  See Fig.~\ref{fig:NCC:compilation} for parameters.
  Because the \Tstraintransfer{} for the \IUfebus{} varies between the
  channels, we compute the regression coefficient with respect to each of the
  six \IUfebus{} channels, which results in 36 data points in the diagrams on
  the right (six \IUfebus{} channels times six channels for the IU under
  investigation).
  The orange horizontal bars mark the median values of the
  six channels for each frequency band.
  The kernel density estimates that define the edges of the violin plot
  illustrate the spread in the distribution of numerical values.}
  \label{fig:STR:compilation}
\end{figure*}
The high waveform similarity allows to measure the \Tstraintransfer{}
\begin{equation}
  \Sstr=\frac{\sum\limits_k \, x_k\,y_k}{\sum\limits_k y^2_k},
  \label{eq:strain:transfer}
\end{equation}
by a linear regression, as introduced by \citet[their eq.~5]{forbriger2025}.
Here $x_k$ is the DAS time series of \Tfiberstrain\
and $y_k$ is the strainmeter time series of \Trockstrain.
We compute the \Tstraintransfer\ with signals recorded after the Mw\,7.4
Taiwan earthquake (see Tab.~\ref{tab:earthquakes}) in the frequency band from
\SI{20}{\milli\hertz} to \SI{200}{\milli\hertz}.
The resulting values are displayed in Fig.~\ref{fig:STR:compilation} (left).
The values for the waveform example in Fig.~\ref{fig:man:Taiwan:OptoDAS} 
are given in Tab.~\ref{tab:OptoDAS:Taiwan:coefficients}.
Like \citet[their Fig.~2]{forbriger2025} we find a substantial scatter between
channels.
Additionally, the values for the \Tstraintransfer\ for different IUs varies
about different average values, with some larger and some smaller than 1.
Nevertheless, the median values for all IUs consistently are close to 
1.1 and thus about \SI{10}{\percent} larger than 1.
This apparent amplification of the strain amplitude with respect to the
strainmeter signals presumably is due to the strain field being altered by
about \SIrange{5}{10}{\percent} in between strainmeters and the \Lvorstollen{}
as discussed above.
Hence, the IUs practically measure \Trockstrain{} amplitude at the location of
the \Lvorstollen{}, which implies a \Tstraintransfer{} for the cemented
fibers.

At higher frequencies, the regression with respect to the strainmeter data is
meaningless because of its parasitic response to vertical ground motion.
For the Albstadt and the Italy earthquakes (see Tab.~\ref{tab:earthquakes}),
we run an intercomparison by computing the regression coefficient (RC)
$\Sstr$, where we choose the \IUfebus\ data for $y_k$ in
eq.~\eqref{eq:strain:transfer}.
We select the \IUfebus\ as a reference, because the scatter of its
\Tstraintransfer\ is least in Fig.~\ref{fig:STR:compilation} (left).
To account for variations in the \IUfebus{} signals between channels, we
compute the RC with respect to each of the six \IUfebus{} channels,
which results in 36 data points.
The resulting values for all three earthquakes are shown in
Fig.~\ref{fig:STR:compilation} (right).
Variations of the median value between IUs and between the earthquakes
are clearly smaller than the scatter. 
For this reason, we are not looking for subtle systematic effects in these data.
\subsection{Coherent background signals}
The subtraction of the signals recorded on the reference coils removes a
substantial amount of coherent noise from the signals recorded in the groove,
as is demonstrated in Fig.~\ref{fig:spec:QuantX:20240326} (left).
Still, the background signal in adjacent channels shows coherent components in
Fig.~\ref{fig:spec:QuantX:20240326} (right), for frequencies at which
\Tfiberstrain\ represents the Rayleigh waves of the marine microseism.
The large values of NCC found in the intercomparison in
Fig.~\ref{fig:NCC:compilation} corroborate this interpretation of high levels
of coherence at large \Trockstrain\ amplitude.

\begin{figure*}
  \begin{center}
  \includegraphics[width=0.7\textwidth]{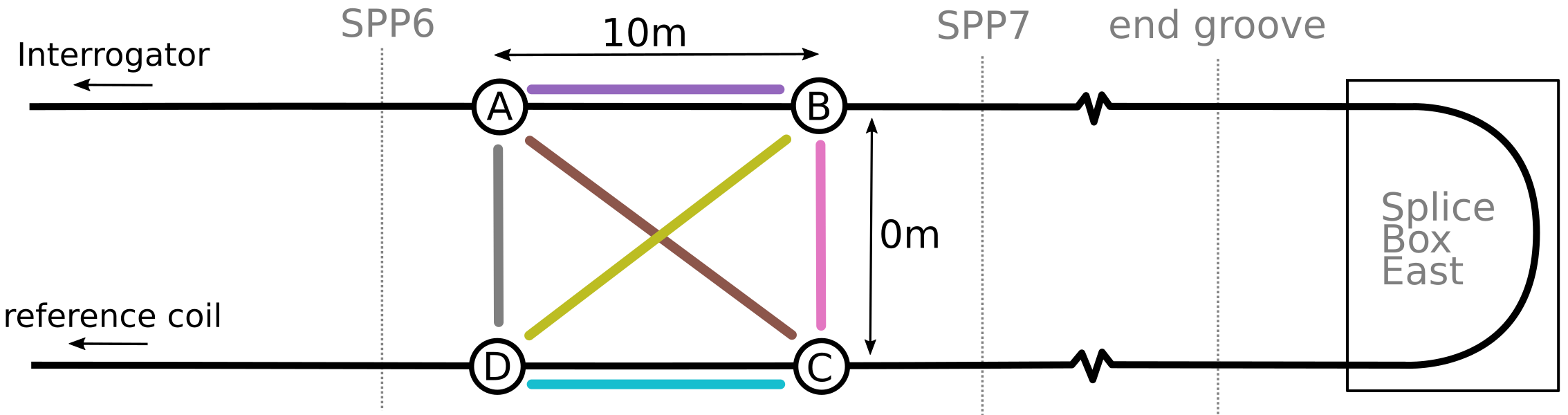}\par
  \bigskip
  \includegraphics[clip,trim=0 0 0 33,width=\textwidth]{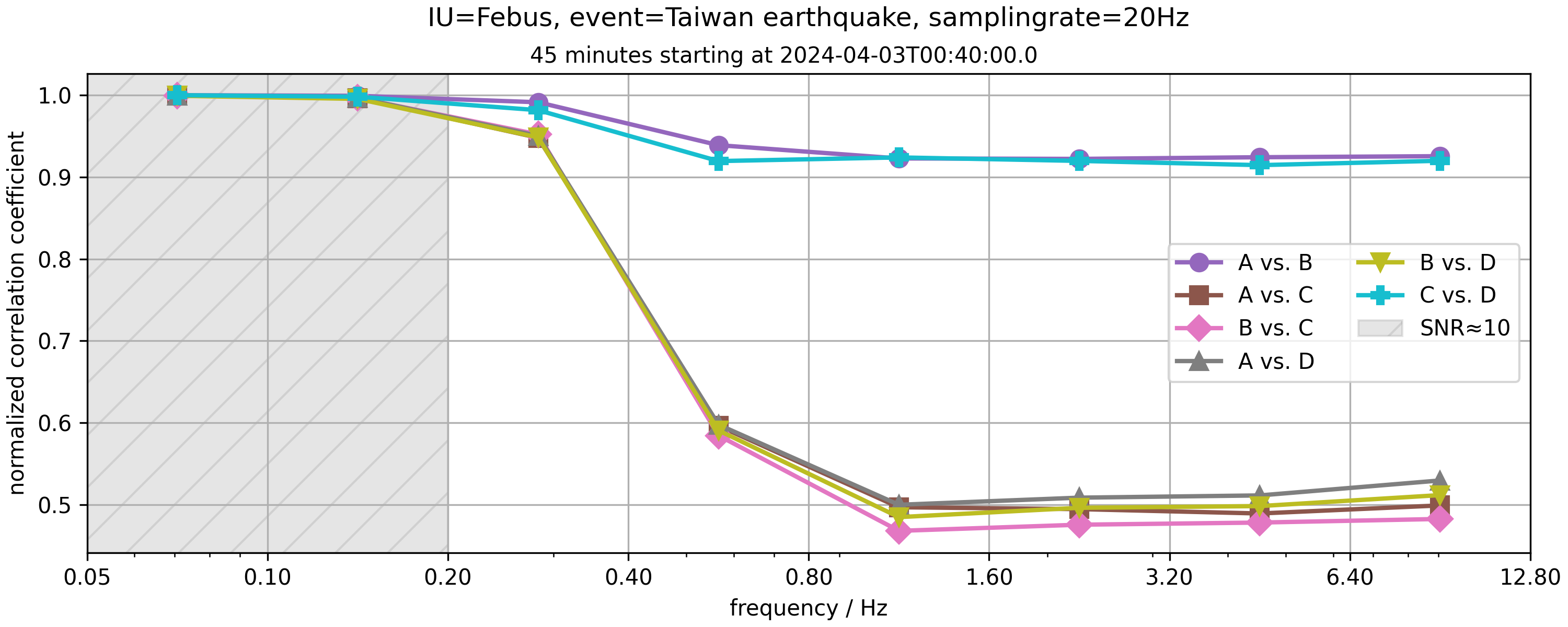}
  \end{center}
  \caption{Signal similarity between adjacent channels.
  Top: Layout of analyzed channels (not to scale).
  Bottom: Signal similarity for pairs of adjacent channels recorded by the
  \IUfebus.
  Data from four locations on the same fiber are selected, where two of them
  (A and B) are in front of the turnaround of the cable in the
  \tunnellocation{Splice Box East} and the two other (C and D) are behind.
  Although the two sections of the cable are right next to each other in the
  groove, A and D as well as B and C need not be exactly adjacent to each
  other due to the constant read-out interval of DAS channels.
  While A and B as well as C and D are at a distance of the read-out interval,
  A and D as well as B and C are considered to be closer than the read-out interval.
  Because the read-out interval is about \SI{10}{\meter} and the gauge length
  is chosen close to \SI{50}{\meter} (Tab.~\ref{tab:IUs}) the pairs A and B as
  well as C and D sample a common section of the same fiber.
  The pairs A and D as well as B and C sample a common section of the rock on
  different fibers.
  The analyses
  signals were recorded during \SI{45}{\minute} after 2024-04-03 00:40~UTC
  (the time window of the wave train from the the Mw\,7.4 Taiwan
  earthquake, see Tab.~\ref{tab:earthquakes}).
  They are compared for different
  frequency bands (tick marks on the abscissa give the upper and lower limit).
  The value of the NCC for channel pairs sketched in the upper diagram
  is displayed in the respective frequency interval in the lower diagram.
  For frequency below \SI{0.2}{\hertz} the earthquake signal dominates the
  recordings and the NCC is large for all of the pairs.
  The signal-to-noise ratio for the earthquake drops with increasing
  frequency.
  Pairs on a common section of the fiber then produce a larger value of NCC.}
  \label{fig:A1R:adjacent}
\end{figure*}

By a frequency-dependent computation of NCC, we investigate how the coherence of
adjacent channels falls off outside the frequency band of large \Trockstrain\
SNR.
As sketched in Fig.~\ref{fig:A1R:adjacent} (top)
we compare adjacent channels (A
with B, as well as C with D) on the same fiber and channels on adjacent fibers
(A with C and D, as well as B with C and D).
Fig.~\ref{fig:A1R:adjacent} (bottom)
shows the corresponding NCC values for the
\IUfebus\ and the Mw\,7.4 Taiwan earthquake (see Tab.~\ref{tab:earthquakes}).
For the frequency bands up to \SI{0.2}{\hertz}, the signal amplitude provides a
high SNR (near 10).
At higher frequencies, there is little energy from the earthquake in the
recording, and NCC values for channels of adjacent fibers drop with increasing
frequency, indicating that there is no common \Trockstrain\ signal picked up.
This is not the case for signals recorded on the same fiber (A vs.\ B and C
vs.\ D), which means that there are similarities in \Tfiberstrain\ though it
does not represent a common \Trockstrain\ signal.
We see similar behavior for signals recorded with the \IUoptodas\ and the
\IUquantx, though not as clear as with the \IUfebus.

\begin{figure*}
  \begin{center}
  \includegraphics[width=0.7\textwidth]{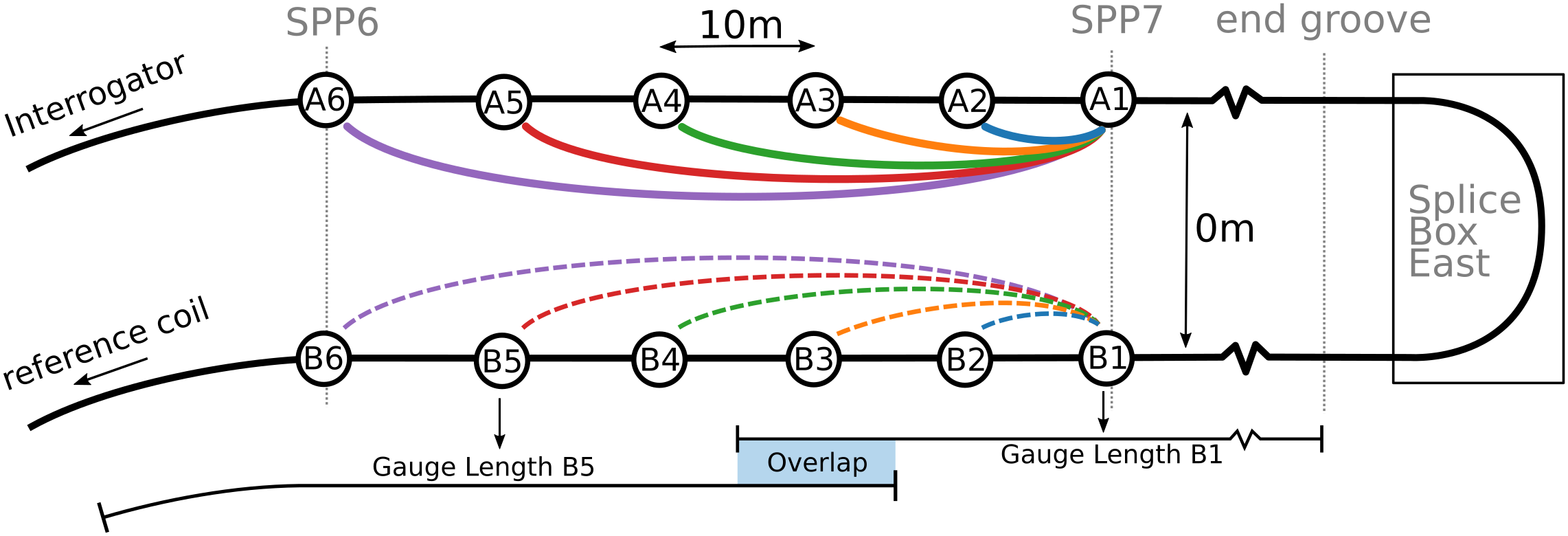}\par
  \bigskip
  \includegraphics[clip,trim=0 0 0 33,width=\textwidth]{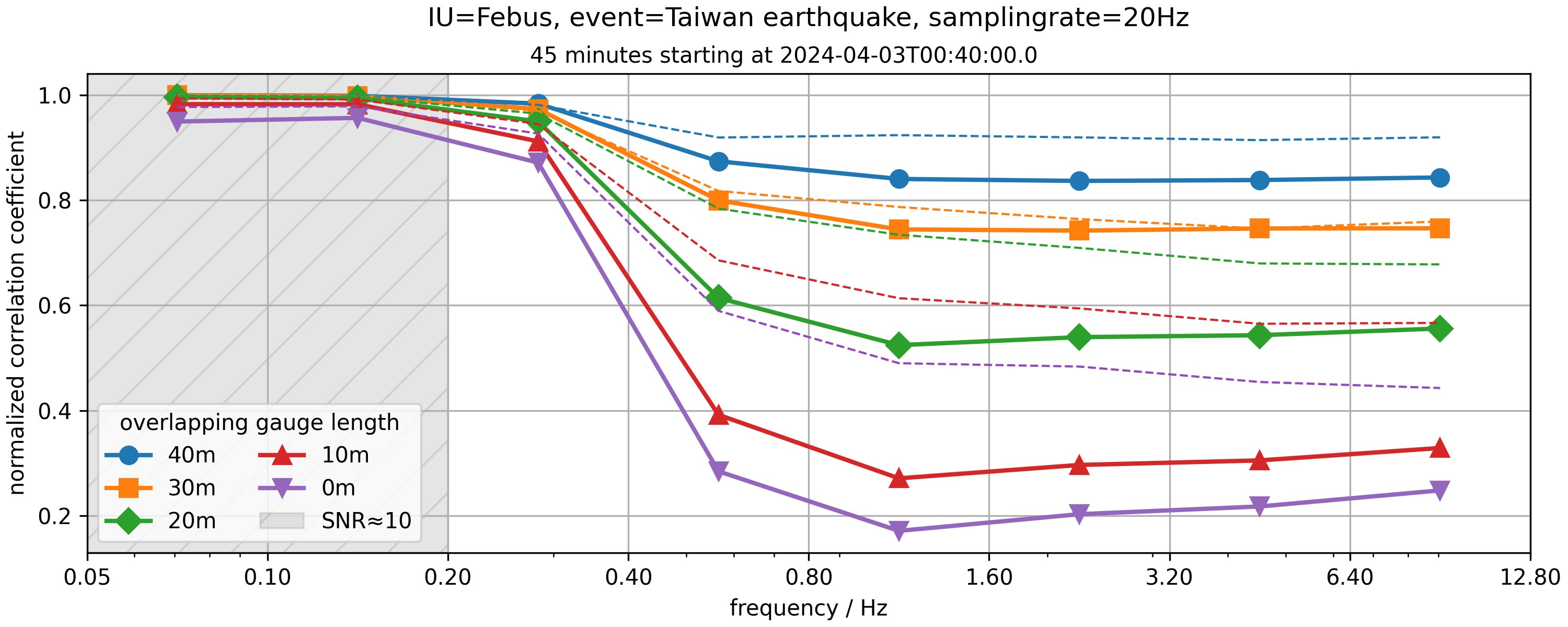}
  \end{center}
  \caption{Signal similarity between overlapping channels.
  Top: Layout of analyzed channels (not to scale).
  Bottom: Signal similarity for pairs of overlapping channels recorded by the
  \IUfebus.
  Channels A1 and B1 are on the same cable at opposite sides of the turnaround
  in the \tunnellocation{Splice Box East} and are selected to be at a
  spatial distance smaller than the read-out interval.
  The two sections of the cable are right next to each other in the
  groove.
  Data from each of them are compared to signals of channels which are at a
  distance increasing stepwise with the read-out interval
  from A2 to A6 and B2 to B6, respectively.
  Because the read-out interval is about \SI{10}{\meter} and the gauge length
  is chosen close to \SI{50}{\meter} (Tab.~\ref{tab:IUs}) these pairs have
  different overlap on the common fiber.
  The time window and the frequency bands are the same as in
  Fig.~\ref{fig:A1R:adjacent}}.
  \label{fig:A1R:overlapping}
\end{figure*}

While the gauge length is about \SI{50}{\meter}, the distance of adjacent
channels on the same fiber is about \SI{10}{\meter}.
Both channels hence sample a common section of \SI{40}{\meter}.
We further investigate this by analyzing channels on the fiber with different
amounts of overlap, as sketched in Fig.~\ref{fig:A1R:overlapping} (top).
The resulting NCC values are displayed in Fig.~\ref{fig:A1R:overlapping}
(bottom) for the \IUfebus.
As expected, the NCC is lower the smaller the common section of the fiber.
However, even for no overlap there remains some coherent signal component.
At the same time NCC slightly drops with decreasing overlap in the frequency
band of high SNR, as one of the channels in the pair leaves the straight
stretch of the \Lvorstollen\ and apparently measures a slightly different
component of the strain tensor.
For the \IUquantx\ and the \IUoptodas\ the drop of NCC with increasing
frequency is even stronger for most channel combinations, meaning that less
coherent signal remains even with a finite overlap.

\end{multicols}
\clearpage
\section{Conclusions}
\begin{multicols}{2}
On a testbed with fiber optic patch cables cemented into the concrete floor of
BFO, we investigate whether the amplitude loss, as indicated by a
\Tstraintransfer\ less than one in previous tests by \citet{forbriger2025},
can be overcome by tighter coupling.
As a reference we take waveforms from the array of calibrated strainmeters at
BFO in the frequency band \SIrange{0.02}{0.2}{\hertz}.
Taking the slight spatial distortion of the strain field due to surface
topography into account, we find a \Tstraintransfer{} of practically 1 for the
cemented fibers.
This result is corroborated by the data from four different IUs.
At higher frequency (\SIrange{0.02}{14}{\hertz}) an intercomparison of data
from different IUs confirms that up to \SI{14}{\hertz} no IU-specific
frequency dependence of \Tstraintransfer{} can be seen in consideration of
scatter between different channels.

For this analysis we use earthquake signals with high waveform similarity
(the normalized correlation typically being $\ge 0.95$, see
Fig.~\ref{fig:NCC:compilation}) between recordings
of DAS IUs and strainmeters or between IUs.
The \Tstraintransfer\ for this reason can be measured in a phase-sensitive way
by the regression coefficient.
Between channels recorded by a single IU, we observe a scatter of about
$\pm\SI{10}{\percent}$ in most cases (Fig.~\ref{fig:STR:compilation}), 
where the high waveform similarity would
suggest a smaller variation in amplitude.
This scatter might be due to the variation of effective \qmarks{gauge length}
caused by the randomness of spatial distribution of backscatterers.
The variation of the median values in between earthquakes and IUs, however, 
typically is less than \SI{5}{\percent}.

Three of the IUs see a background signal which is coherent between all
channels along the fiber route, presumably so-called common mode laser noise
\citep[][their section 2.6 Optical Noise]{lindsey2020}.
We improve the signal-to-noise ratio by up to \SI{20}{\decibel} by
subtracting the average signal recorded on \qmarks{reference coils}.
In that way, we reduce the background noise level in the best cases to
$\SI{100}{\pico\strain}$ at $\SI{0.1}{\hertz}$ and $\SI{5}{\pico\strain}$ at
$\SI{1}{\hertz}$ in a bandwidth of 1/6 decade
(see Fig.~\ref{fig:spec:QuantX:20240326}).
Three of the IUs detect the marine microseisms at times of moderate amplitude.
The remaining background signal is largely incoherent for channels on
co-located fibers.
The coherence of the background between neighboring channels increases with increasing overlap of their gauge length.

The background level in the analyses is most likely controlled by various
factors of the installation, like thermal effects on the reference coils
or vibration isolation being different for the different IUs.
The background signal level hence not necessarily represents a property of the
respective IU alone.

The procedure of noise reduction makes the reference coils a critical
component, which shall be protected from local disturbances like dripping
water or strong temperature fluctuations.
The remaining background signal is so low, that it might be due to
temperature fluctuations of the fiber (possibly on the reference coil)
at a level of a few \si{\micro\kelvin}, given that the temperature coefficient of the
fiber refractive index is on the order of $10^{-5}\,\si{\per\kelvin}$.

\end{multicols}
\clearpage
\begin{multicols}{2}
  \raggedcolumns
\section*{Data and Resources}
  All data were recorded locally at
  \citet[][\doiurl{10.5880/BFO}]{bfo1971}.
  Strainmeter and seismometer data are available through
  data centers at the \citet[][\doiurl{10.7914/SN/II}]{sio1986}
  and the \citet[][\doiurl{10.25928/MBX6-HR74}]{bgr1976}, respectively.
  Data and software used in this paper are provided for download by
  \citet[][\doiurl{10.35097/gj4yg6478gtesxdw}]{forbriger2026data}.
  Data analysis was carried out with ObsPy
  \citep[][\doiurl{10.5281/zenodo.6327346}]{obspy2022}.
\section*{Supplementary material}
  Supplementary material is provided in the second part of this document 
  (pages \pageref{sec:supp:introduction} -- \pageref{fig:MM:PostSwap:Treble}).
  It provides time series examples for
all IUs and all earthquakes listed in Table~\ref{tab:earthquakes}.
For all IUs, a spectral analysis of the background signal in a time window
  before
and after swapping the fiber routes is presented, as well as a waveform example
with and without subtraction of the reference coil signal.
IU configuration parameters are given in tables.
The supplementary text
further discusses timing accuracy, the means used to locate the
individual channels in the mine, and the accuracy by which this can be done.
This is complemented by a discussion of scaling factors potentially affecting
the strain amplitude presented by the IUs.
For the marks SPP1 to SPP7 in Fig.~\ref{fig:floor:map:test:bed} their location
is given in terms of offset to the IU along the fiber route 
as well as their UTM coordinates.

\section*{Author's contributions}
According to the CRediT Taxonomy:
Conceptualization: all authors;
Data curation: TF, LH;
Formal analysis: TF, FM, LH, HX;
Investigation: TF, LH, RWS, AR, HX;
Methodology: TF;
Project administration: TF, RWS, LH;
Resources: RWS, AR, TF, LH, VRT, HX, AS;
Software: TF, LH;
Visualization: TF, FM;
Writing – original draft: TF, VRT, LH, FM;
Writing – review \& editing: all authors 

\section*{Acknowledgments}
We thank Peter Duffner for taking care of the BFO strainmeters and their
  calibration devices.
We are grateful to Walter Zürn for many fruitful discussions and his
  continuing support of the operation of the Invar-wire strainmeters.
Felix Bögelspacher, Jérôme Azzola, and Nino Krämer helped with the
installation of the testbed.
We are grateful to the teams of Alcatel Submarine Networks (ASN), Terra15,
LUNA OptaSense, and Febus for helpful discussions of our observations.
\end{multicols}
\printbibliography
\clearpage
%
%
%
%
%
\appendix
\setcounter{figure}{0}
\setcounter{table}{0}
\setcounter{section}{0}
\setcounter{equation}{0}
\renewcommand{\thefigure}{S\arabic{figure}}
\renewcommand{\thetable}{S\arabic{table}}
\renewcommand{\thesection}{S\arabic{section}}
\renewcommand{\theequation}{S\arabic{equation}}
\newcommand{\FigAlbstadtEQ}[5]{
\begin{figure}
  {\includegraphics[trim=0 0 0 58,clip,width=\textwidth]{#1_20240322AlbstadtEQ_BW_fiberstrain#4}}
  \caption{Waveform comparison for the Albstadt earthquake (see
  Table~\ref{tab:earthquakes}) recorded with the #2.
  All signals are band-passed to frequencies from \SI{3}{\hertz} to
  \SI{18}{\hertz}.
  Optical channel offsets (all traces located in the \Lvorstollen)
  are given in the legend. Fiber extension is displayed, no correction for
  \Tstraintransfer\ is applied. #5}
  \label{#3}
\end{figure}
}
\newcommand{\FigItalyEQ}[5]{
\begin{figure}
  {\includegraphics[trim=0 0 0 58,clip,width=\textwidth]{#1_20240327ItalyEQ_BW_fiberstrain#4}}
  \caption{Waveform comparison for the Italy earthquake (see
  Table~\ref{tab:earthquakes}) recorded with the #2.
  All signals are band-passed to frequencies from \SI{0.2}{\hertz} to
  \SI{8}{\hertz}.
  Optical channel offsets (all traces located in the \Lvorstollen)
  are given in the legend. Fiber extension is displayed, no correction for
  \Tstraintransfer\ is applied. #5}
  \label{#3}
\end{figure}
}

\newcommand{\nonrcomment}{No noise reduction (based on the reference coils) is
applied.}
\newcommand{\nrcomment}{The average signal recorded on the reference coils is
subtracted.}
\newcommand{\FigTaiwanEQ}[5]{
\begin{figure}
  \includegraphics[trim=0 0 0 60,clip,width=\textwidth]{#1_20240403TaiwanEQ_SW_adjusted}
  \includegraphics[trim=0 0 0 60,clip,width=\textwidth]{#1_20240403TaiwanEQ_SW_adjusted_nonr}
  \caption{Waveform comparison for the Taiwan earthquake (see
  Table~\ref{tab:earthquakes}) recorded with the #2.
  The traces are vertically shifted for better visibility of details.
  The S-phase arrives at about 2024-04-03 00:21 UTC.
  The P-phase (not shown) does not exceed the noise level.
  Rayleigh waves arrive after \SI{1300}{\second} on the time scale.
  All signals are band-passed to frequencies from \SI{20}{\milli\hertz} to
  \SI{200}{\milli\hertz}.
  The top trace is the signal from the BFO strainmeters in \azimuth{90}
  azimuth.
  The other six traces are recorded by the #2 in the \Lvorstollen{}.
  Optical channel offset and amplitude scaling factors are given in the
  legend, where the scaling factors are the reciprocal \Tstraintransfer. 
  Upper diagram: \nrcomment
  Lower diagram: \nonrcomment}
  \label{#3}
\end{figure}
}
\newcommand{\FigMMrms}[3]{
\begin{figure}
  {\includegraphics[clip,trim=0 0 0 34,height=63mm]{#1_rms}}
  \hfill
  {\includegraphics[clip,trim=0 0 0 34,height=63mm]{#1_coherence}}
  \caption{#2}  
  \label{#3}
\end{figure}
}
\newcommand{\MMcaption}[3]{%
    Analysis of the background signal on #1
    for recordings of the #2:
    rms amplitude in 1/6 decade \parencite[left, see section
    4.1 by][]{forbrigerPSILN2023} and magnitude-squared coherence 
    \citep[their eq.~2]{carter1973} with respect to the BFO strainmeter data
    (right).
    The corners of the applied bandpass are at \SI{20}{\milli\hertz} and
    \SI{15}{\hertz} and thus lie outside the displayed range.
    We compare signals from six channels (indicated by their offsets) to the
    \Trockstrain\ in \azimuth{90} as obtained from the BFO strainmeter array
    (purple curve).
    The curve for the latter is truncated at \SI{0.5}{\hertz}, because
    of the parasitic sensitivity to vertical ground motion at higher
    frequencies.
    The dashed lines in the left diagram show the signal level of the raw DAS
    data.
    The solid lines are the levels after subtracting the average of the signal
    on the reference coil in the time domain.
    Signal amplitudes are not further scaled to correct for \Tstraintransfer\
    (factor: 1.0).}
\newcommand{\FigPreSwapMMrms}[5]{
  \FigMMrms{#1_#2_20240308PreSwapBackground_BP20-15000_MM}{\MMcaption{2024-03-08
  \SIrange{1}{2}{\UTC}}{#5}}{#4}
}
\newcommand{\FigPostSwapMMrms}[5]{
  \FigMMrms{#1_#2_20240326PostSwapBackground_BP20-15000_MM}{\MMcaption{2024-03-26 
  \SIrange{1}{2}{\UTC}}{#5}}{#4}
}
\title{Supplemental material to\\
\emph{\mytitle}}
\author{Thomas Forbriger,
Felix Münch,
Laura Hillmann,
Ver\'onica Rodr\'iguez Tribaldos,\\
Rudolf Widmer-Schnidrig,
Han Xiao,
Andreas Rietbrock,
Angelo Strollo,
Philippe Jousset}
\date{}
\maketitle

\section{Introduction}
\label{sec:supp:introduction}
\begin{multicols}{2}
This supplement adds some technical details regarding the location and time
accuracy as well as signal amplitude scaling.
Results for all the analyses discussed in the main manuscript are displayed
her for each of the interrogator units (IUs) separately.
Though we do not intend to rank the IUs in any way.
Please consider the respective comment in section
\ref{sec:IUs}~\qmarks{\nameref{sec:IUs}}.

\end{multicols}
\section{Channel locations}
\begin{multicols}{2}
Individual sensors, such as seismometers, have a known location that is
selected and georeferenced during installation. 
In contrast, fiber optic cables are a spatially extended sensor that, in our
case, was folded multiple times over a distance of more than
\SI{1}{\kilo\meter} in a narrow tunnel.
Specific DAS channels located in the \Lvorstollen, for example, are identified
by their distance to the IU.
Establishing this connection between linear distance and specific location in
the case of a cable with multiple folds is a challenge in itself.

The most reliable basis for this are so-called tap-tests.
Vibrations are excited at well defined locations in the tunnel by tapping on
the ground (we stamped a foot on the concrete floor next to the mark). In order to improve the location accuracy as much as possible, tap tests were recorded using a short GL and small channel interval
 (both typical \SI{1}{\meter}) on 2024-02-16, prior to the actual experiment
which had a larger GL and channel interval. 
or \SI{2}{\meter}). The tap test signals were identified in the data and the linear distance to the IU was determined. 
Despite the small GL and channel interval during the tap test, there remains a
significant inaccuracy for at least three reasons:
1) the vibrations propagate as elastic waves and are present on several
channels,
2) optical fading modulates the signal amplitudes between channels, such that
the channel of largest amplitude might not be the closest to the tap, and
3) signal to noise ratio at small GL typically is reduced and some of the tap
signals might be missed.
For this reason, we have cross-checked the consistency of the tap-test results
against distance marks printed on the cable and against OTDR measurements,
which allowed us to identify some (not all) of the fusion splices in the
splice boxes.
The marks printed on the cables alone cannot solve the problem, because the
concatenation of several cables accumulates the inaccuracy of pigtail lengths
and because the Febus \IUfebus\ adds an internal lead-in of approximately
\SI{25}{\meter} to the total fiber route (where this is not precisely
specified in the manual).

\sisetup{separate-uncertainty}%
We at least know that the section from SPP2 to SPP7 (see
Fig.~\ref{fig:floor:map:test:bed} in the main text) is sampled eight times.
By taking the distance to SPP7 and averaging over the eight cable sections, we
increase the precision of the optical distances between the marks.
The relative distances vary by less than \SI{1}{\meter} between the eight
sections in most cases.
Their inaccuracy increases with distance to SPP7 and can become almost
\SI{5}{\meter} near SPP2 in some cases.
We find the distance between SPP7 in front and behind the splice box east to
be \SI{66.7(7)}{\meter} very consistently for all four pairs.
This way, we compose Table~\ref{tab:fiber:route:distances} for the linear
distances between the IUs and the marked locations in the mine.
The GL of \SI{50}{\meter} in the actual experiment is so large that practically
every channel along the groove, even those near the splice box east, picks up
rock strain signals during earthquakes.
By comparison of signal levels along the fiber route, we convince ourselves
that the splice box section, as expected from the values in
Table~\ref{tab:fiber:route:distances}, becomes apparent by reduced coupling.

The Table~\ref{tab:fiber:route:distances} is complemented by
Table~\ref{tab:utm:coordinates} which specifies the absolute coordinates of
marked points as derived from geodetic surveying.

\end{multicols}
\section{Timing accuracy}
\begin{multicols}{2}
As discussed in the main text,
timing was provided to the \IUquantx\ and the \IUtreble\ through GPS antennas.
The \IUfebus\ also used a GPS signal provided through an optical link
(Meinberg GOAL: GPS Optical Antenna Link).
For the \IUoptodas{}, GPS reception from the optical link was unsuccessful and
time was synchronized to a 1\,PPS signal and the \SI{10}{\mega\hertz} signal from
the GOAL.

We checked the timing accuracy by cross-correlating signals of co-located
channels of different IUs for the short period signals of the earthquakes near
Albstadt and Italy (see Table~\ref{tab:earthquakes} in the main text).
We find timing offsets between \SI{6}{\milli\second} and
\SI{40}{\milli\second}, which are not exactly stable over a longer period of
several days.
While the origin of these offsets is not known, they do not affect the results
for the low-frequency investigations (marine microseism background and the
Hualien City earthquake).
For the short period earthquakes, the time offsets still are smaller than the
shortest investigated signal period band of \SIrange{55}{333}{\milli\second}
for the Albstadt earthquake.
To reduce their potential impact, we have time shifted the signals during
analysis by the offset measured with cross-correlation in order to align the
compared signals with each other.

The finite speed of wave propagation along the fiber can add additional signal
delays, in particular for surface waves incoming from the east. For the
Albstadt earthquake (BAZ approximately to the east), we find an average delay
of \SI{2.5}{\milli\second} over the \SI{10}{\meter} channel spacing.
This is consistent with a phase velocity of \SI{2.5}{\kilo\meter\per\second},
which is in the expected order or magnitude for surface waves in the local
crustal structure.

\end{multicols}
\section{Scaling to strain}
\begin{multicols}{2}
DAS interrogators in first place measure a change of the phase of the light
pulse returned from the fiber with respect to some reference.
Factors controlling the conversion of phase change $\Delta\Phi$ to fiber strain
$\epsilon$, which caused the phase change, are
the optical wavelength $\lambda$ in vacuum of the laser,
the refractive index $n$ of the fiber,
the opto elastic factor $\xi$ of the fiber,
and the gauge length $G$.
The opto elastic factor $\xi$ accounts for the change of refractive index, if
the fiber gets strained, as described by \citet[his
eq.~6.6 in his chapter 6.2]{hartog2017introduction}.
For IUs which convert the phase change $\Delta\Phi$ directly to fiber strain
\begin{equation}
  \epsilon=\frac{\lambda\,\Delta\Phi}{4\pi\,n\,G\,\xi}
\end{equation}
the respective equation is given in the document by
\citet[their EQ~10]{seafom2024}.

$\lambda$ is an instrument property of the IU and $G$ is set by the
configuration for the experiment.
$n$ and $\xi$, however, are properties of the fiber being in use.
Their exact values are not available to the IU and some of the IUs use
standard values for fused silica without allowing to adjust them in the
configuration.
In fact, the fiber manufacturers commonly do not specify the exact values for
their fibers.

During the fabrication process of the fiber $n$ is adjusted by dopants
\citep[their section 2.2.3 Fiber Fabrication]{sillard2017}.
This at least is needed to separate the cladding from the core to provide
total reflection but does not change $n$ by more than \SI{1}{\percent}.
In graded fibers $n$ varies continuously with the radius
\citep[their Figure~9]{horiguchi1980}.
Further, the index of refraction is dispersive \citep{lee2016}.
At $\lambda=\SI{1550}{\nano\meter}$, the wavelength used by the IUs in the
current experiment, the value for the doped fiber typically falls in the
range of $n=\text{\numrange{1.46}{1.47}}$.
In the absence of more accurate values a choice of
$n=1.47$ appears reasonable.

\citet{abe1995} have explored the range for the reciprocal of
$\xi$ at large strain for various specimens of fibers and find values of
$\xi^{-1}=\text{\numrange{1.26}{1.30}}$
for the smaller strain values and $\lambda=\SI{1550}{\nano\meter}$.
This range corresponds to $\xi=\text{\numrange{0.77}{0.794}}$.
$\xi=0.78$, a value consistent with the elastic parameters of silica
\citep[their eq.~1]{giallorenzi1982}, appears to be a reasonable choice if
no information regarding the actual properties of the fiber are available.

The temperature sensitivity of the refractive index lies in the order of
magnitude of $\SI{1.e-5}{\per\kelvin}$ and is large enough to cause an
apparent strain signal due to temperature variation.
It is however negligible with respect to its consequences for signal amplitude
scaling.

\end{multicols}
\section{Interrogators}
\begin{multicols}{2}
\label{sec:IUs}
In the following section we list selected recording parameters for each of the
IUs.
We supplement the figures from the main manuscript with diagrams for each of
the IUs.
Before doing so, we repeat a statement from the mean manuscript, regarding
intentions:

We use four IUs in order to corroborate the properties of the testbed and the
strain transfer rate in particular, by validation with IUs which implement
different techniques.
We do not intend to rank the IUs in any way.
The recording parameters and analyzed frequency bands for the four IUs have
been chosen as similar as possible, which in turn means, that these parameter
might not be optimal for each of the IUs.
In most of the data we find details, which we still do not fully understand.
Some of them might be due to the recording conditions and be caused by
water occasionally dripping from the mine ceiling onto the fibers, 
by temperature fluctuations, or by the different
installation conditions for the IUs themselves.
These details need not represent a property of the respective IU.
Where disturbances are obvious, we have not used the affected time windows.
In the following, we discuss the gross features of the DAS data, in particular
with respect to the coupling of the fiber to the rock and the potential of this technology to record rock strain.
We will mention similarities and differences between data from different IUs,
but like to remind the reader that these might be specific to the
particular setup.

The effect of subtracting the average of signals recorded on the reference
coils is demonstrated by showing strain seismograms for the Taiwan earthquake
with and without this method, which reduced coherent noise from the
recordings.
This measure of noise reduction is applied to the fiberstrain seismograms of
the \IUfebus{}, \IUquantx{}, and \IUoptodas{} but not for the \IUtreble{} for
the Albstadt and the Italy earthquake.
For background levels of rms-amplitude both versions are shown.
See the figure captions for details.
\end{multicols}

\begin{sidewaystable}
  \caption{Distances at which marked locations along the fiber routes are
  found in the DAS data (see Fig.~\ref{fig:floor:map:test:bed} in the main
  text).
  Fiber routes ec1 and ec2 start in the electronics vault and routes lab3 and
  lab4 start in the laboratory building.
  On 2024-03-19 the routes were swapped pairwise in order to rule out that
  specific properties of data and noise levels are due to one of the
  cables.
  The Febus \IUfebus\ adds an internal lead-in of approximately
  \SI{25}{\meter} to the total fiber route.}
  \label{tab:fiber:route:distances}
\newcolumntype{R}{>{\hfill}p{0.13\textwidth}}
\begin{tabular}{l*{6}{R}}
\toprule
           & & & & &OptaSense \IUquantx &Terra15 \IUtreble \\
           &\multicolumn{2}{c}{Febus \IUfebus}
           &\multicolumn{2}{c}{ASN \IUoptodas}
           &before 2024-03-19 &before 2024-03-19\\
           \cmidrule(lr){2-3}\cmidrule(lr){4-5}
           \cmidrule(lr){6-6}\cmidrule(l){7-7}
           & & & & 
           &Terra15 \IUtreble
           &OptaSense \IUquantx \\
           &before 2024-03-19 & after 2024-03-19
           &before 2024-03-19 & after 2024-03-19
           &after 2024-03-19
           &after 2024-03-19\\
mark       & route ec1 & route ec2 &  route ec2 & route ec1 &
  route lab3     & route lab4 \\
           \cmidrule(r){1-1}
           \cmidrule(lr){2-2}
           \cmidrule(lr){3-3}
           \cmidrule(lr){4-4}
           \cmidrule(lr){5-5}
           \cmidrule(lr){6-6}
           \cmidrule(l){7-7}
SPP1       & 442.7 m &  459.6 m &  434.6 m & 417.7 m &  926.7 m &  930.5 m \\
SPP2       & 495.1 m &  512.0 m &  487.0 m & 470.1 m &  979.1 m &  982.9 m \\
SPP3       & 519.5 m &  536.4 m &  511.4 m & 494.5 m & 1003.5 m & 1007.3 m \\
SPP4       & 551.6 m &  568.5 m &  543.5 m & 526.6 m & 1035.6 m & 1039.4 m \\
SPP5       & 577.3 m &  594.2 m &  569.2 m & 552.3 m & 1061.3 m & 1065.1 m \\
SPP6       & 591.9 m &  608.8 m &  583.8 m & 566.9 m & 1075.9 m & 1079.7 m \\
SPP7       & 640.2 m &  657.1 m &  632.1 m & 615.2 m & 1124.2 m & 1128.0 m \\
           \cmidrule(r){1-1}
           \cmidrule(lr){2-2}
           \cmidrule(lr){3-3}
           \cmidrule(lr){4-4}
           \cmidrule(lr){5-5}
           \cmidrule(lr){6-6}
           \cmidrule(l){7-7}
SPP7       & 706.9 m &  723.8 m &  698.8 m & 681.9 m & 1190.9 m & 1194.7 m \\
SPP6       & 755.2 m &  772.1 m &  747.1 m & 730.2 m & 1239.2 m & 1243.0 m \\
SPP5       & 769.8 m &  786.7 m &  761.7 m & 744.8 m & 1253.8 m & 1257.6 m \\
SPP4       & 795.5 m &  812.4 m &  787.4 m & 770.5 m & 1279.5 m & 1283.3 m \\
SPP3       & 827.6 m &  844.5 m &  819.5 m & 802.6 m & 1311.6 m & 1315.4 m \\
SPP2       & 852.0 m &  868.9 m &  843.9 m & 827.0 m & 1336.0 m & 1339.8 m \\
SPP1       & 904.4 m &  921.3 m &  896.3 m & 879.4 m & 1388.4 m & 1392.2 m \\
           \cmidrule(r){1-1}
           \cmidrule(lr){2-2}
           \cmidrule(lr){3-3}
           \cmidrule(lr){4-4}
           \cmidrule(lr){5-5}
           \cmidrule(lr){6-6}
           \cmidrule(l){7-7}
Start coil & 947.1 m &  954.0 m &  929.0 m & 922.1 m & 1426.1 m & 1419.9 m \\
End coil   &1740.1 m & 1188.0 m & 1163.0 m &1715.1 m & 1636.1 m & 1625.9 m \\
\bottomrule
\end{tabular}
\end{sidewaystable}

\clearpage
\begin{table}
  \caption{UTM32N coordinates of marks in the gallery (see
  Fig.~\ref{fig:floor:map:test:bed} in the main text).}
  \label{tab:utm:coordinates}
  \begin{center}
    \begin{tabular}{lll}
      \toprule
      & easting & northing\\
      \cmidrule(rl){2-2}  \cmidrule(l){3-3}
      SPP1  &   449858.53 & 5353094.29 \\
      SPP2  &   449907.03 & 5353113.53 \\
      SPP3  &   449931.23 & 5353119.73 \\
      SPP4  &   449951.44 & 5353144.35 \\
      SPP5  &   449972.55 & 5353157.78 \\
      SPP6  &   449988.29 & 5353160.32 \\
      SPP7  &   450037.14 & 5353159.88 \\
      SPP8  &   450067.41 & 5353160.58 \\
      \bottomrule
    \end{tabular}
  \end{center}
\end{table}
\subsection{Febus \IUfebus}
\begin{table}[h]
  \caption{Parameters used by the Febus \IUfebus\ interrogator when processing
  the data in the unit.}
  \label{tab:febus:parameters}
  \begin{center}
    \begin{tabular}{rl}
      \toprule
      Manufacturer: & Febus \\
      Model: & A1-R\\
      Firmware version: & version 2.2.2\\
      Optical wavelength: & 1550\,nm\\ %
      Refractive index of the fiber: & 1.5 \\ %
      Opto-elastic correction factor: & 1\\ %
      Fiber length: & 1200\,m\\
      Pulse width: & 20\,m\\
      Pulse rate frequency: & 10\,kHz\\
      Block rate: & 1\,Hz\\
      Amplifier power: & 20\,dBm\\
      Sampling resolution: & 80\,cm\\
      Gauge length (GL): & 50\,m\\
      Derivation time (DT): & 20\,ms \\
      Spatial sampling intervalcing: & 9.6\,m\\
      Temporal sampling intervalrval: & 0.005\,s\\
      Downsampling: & with decimation filter\\
      Primarily recorded quantity: & strain-rate\\
      \bottomrule
    \end{tabular}
  \end{center}
\end{table}
\FigTaiwanEQ{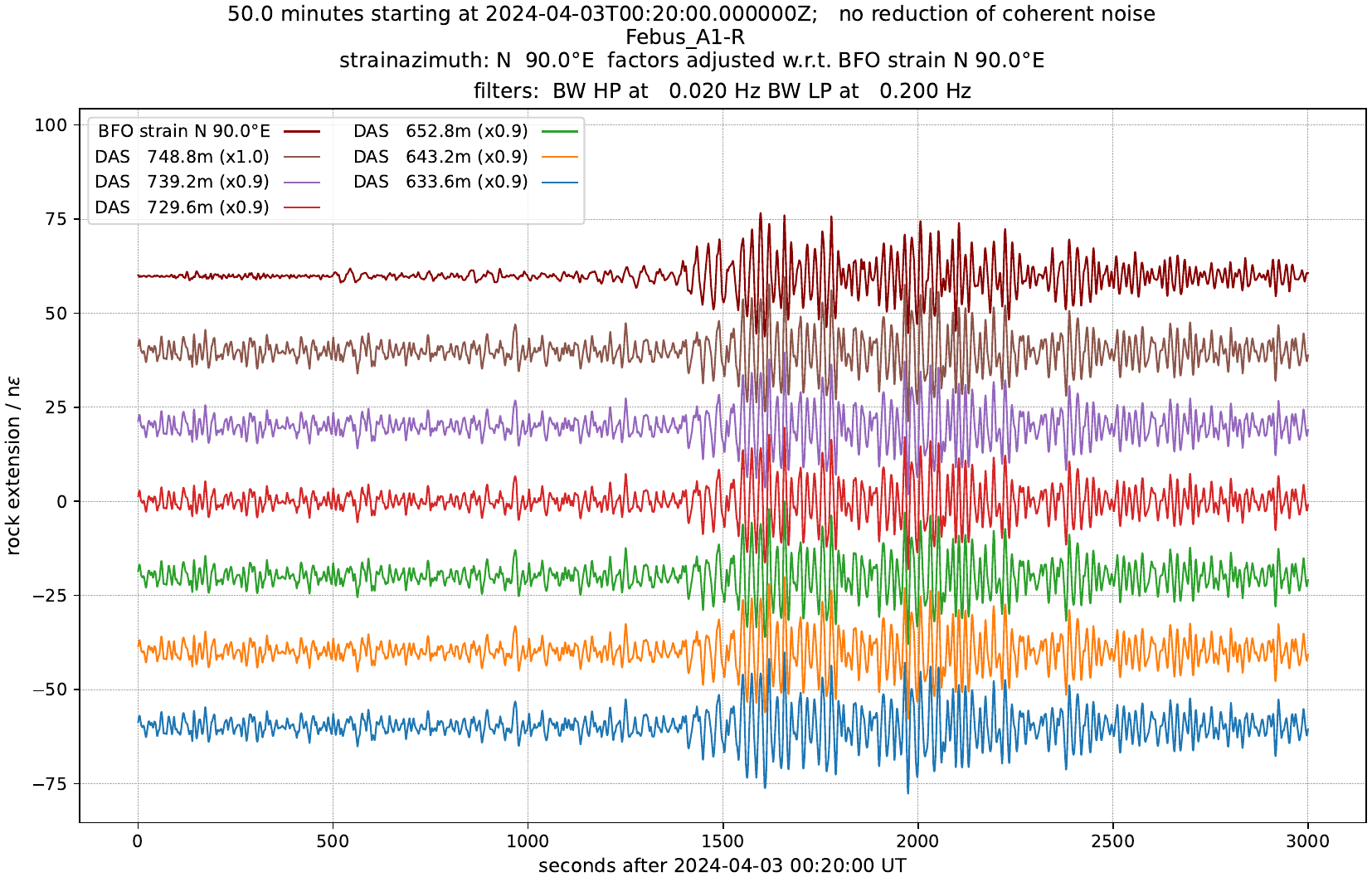}{\IUfebus}{fig:Taiwan:Febus}{}{}

\FigAlbstadtEQ{A1R_Febus_A1-R}{\IUfebus}{fig:Albstadt:Febus}{}{\nrcomment}
\FigItalyEQ{A1R_Febus_A1-R}{\IUfebus}{fig:Italy:Febus}{}{\nrcomment}
\FigPreSwapMMrms{A1R}{Febus_A1-R}{\IUfebus}{fig:MM:PreSwap:Febus}{Febus \IUfebus}
\FigPostSwapMMrms{A1R}{Febus_A1-R}{\IUfebus}{fig:MM:PostSwap:Febus}{Febus \IUfebus}

\clearpage
\subsection{OptaSense \IUquantx}
\begin{table}[h]
  \caption{Parameters used by the OptaSense \IUquantx\ interrogator when
  processing the data in the unit. 
  The values are given with the number of digits as specified in the data file
  headers.}
  \label{tab:quantx:parameters}
  \begin{center}
    \begin{tabular}{rl}
      \toprule
      Manufacturer: & LUNA OptaSense \\
      Model: & QuantX\\
      Firmware Version: & Light Acquisition 5.11.0\\
      Data type: & Diversity Processed Phase XY Dual Pulse Balanced OCP 100\\
      Optical wavelength: & 1550\,nm \\
      Refractive index of the fiber: & 1.4682 \\ %
      Opto-elastic correction factor: & 0.78\\ %
      Pulse width: & 250\,ns \\
      Pulse rate: & 10\,kHz\\
      Gauge length (GL): & 51.048\,m \\
      Spatial sampling inteval: & 10.209524154663086\,m\\
      Temporal sampling interval: & 0.005\,s\\
      Primarily recorded quantity: & strain\\
      \bottomrule
    \end{tabular}
  \end{center}
\end{table}

\FigTaiwanEQ{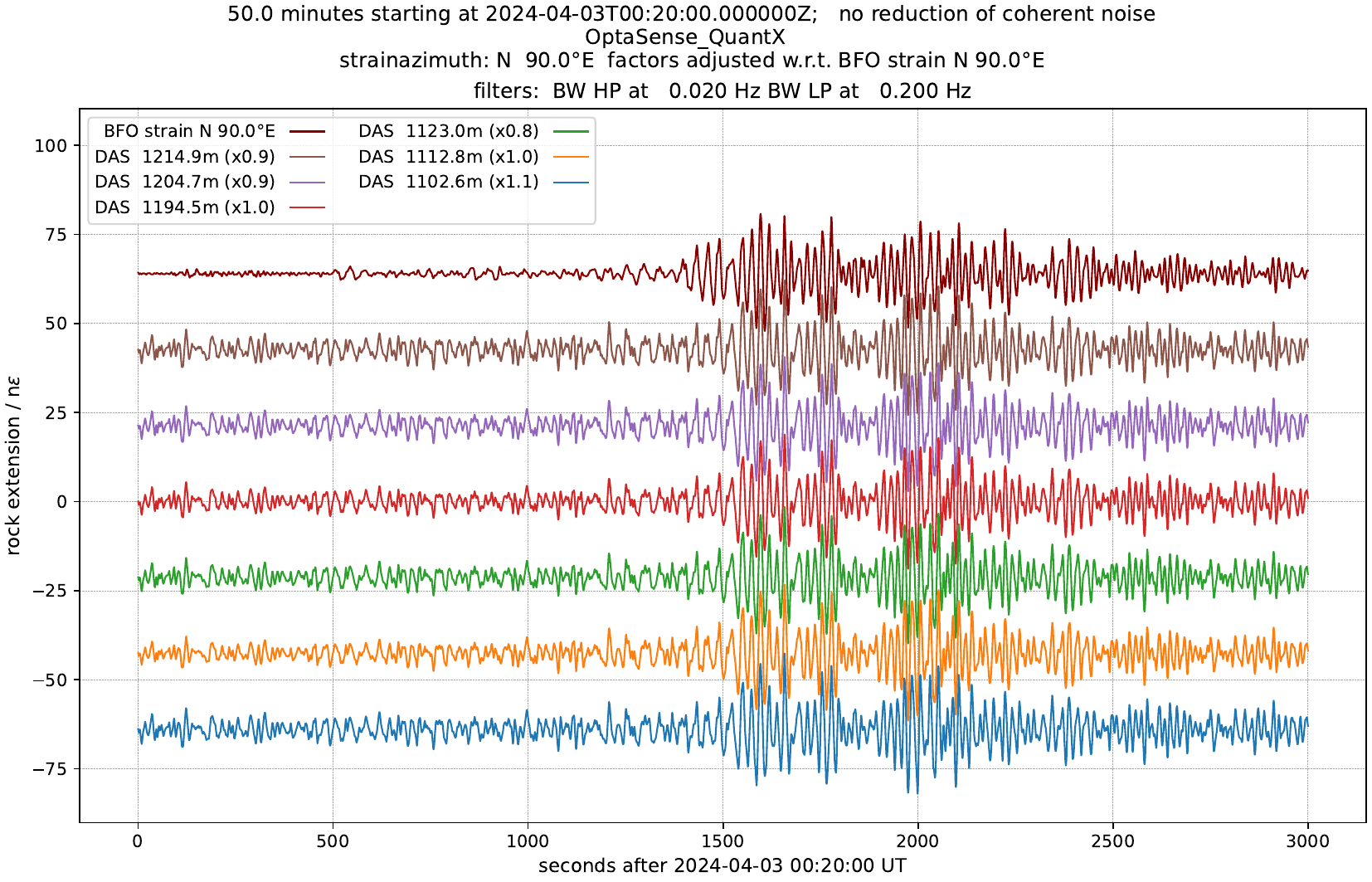}{\IUquantx}{fig:Taiwan:QuantX}{}{}

\FigAlbstadtEQ{QuantX_OptaSense_QuantX}{\IUquantx}{fig:Albstadt:QuantX}{}{\nrcomment}
\FigItalyEQ{QuantX_OptaSense_QuantX}{\IUquantx}{fig:Italy:QuantX}{}{\nrcomment}
\FigPreSwapMMrms{QuantX}{OptaSense_QuantX}{\IUquantx}{fig:MM:PreSwap:Quantx}{OptaSense
\IUquantx}
\FigPostSwapMMrms{QuantX}{OptaSense_QuantX}{\IUquantx}{fig:MM:PostSwap:Quantx}{OptaSense
\IUquantx}

\clearpage
\subsection{ASN \IUoptodas}
\begin{table}[h]
  \caption{Parameters used by the ASN \IUoptodas\ interrogator when
  processing the data in the unit.
  The values are given with the number of digits as specified in the data file
  headers.}
  \label{tab:optodas:parameters}
  \begin{center}
    \begin{tabular}{rl}
      \toprule
      Manufacturer: & Alcatel Submarine Networks (ASN) \\
      Model: & OptoDAS C01-S\\
      Firmware Version: & DasControl Version: 2403011412 \\
      Optical wavelength: & 1536.61\,nm \\ %
      Refractive index of the fiber: & 1.4677 \\ %
      Opto-elastic correction factor: & 0.78\\ %
      Sweep bandwidth: & \SI{80}{\mega\hertz} \\
      Length of compressed chirp pulse: & $\approx$ \SI{12.5}{\nano\second} \\
      Pulse rate: & 50\,kHz\\
      Gauge length (GL): & 40.852\,m\\
      Spatial sampling interval: & 10.213\,m\\
      Temporal sampling interval: & 0.004\,s\\
      Primarily recorded quantity: & strain-rate\\
      \bottomrule
    \end{tabular}
  \end{center}
\end{table}

\FigTaiwanEQ{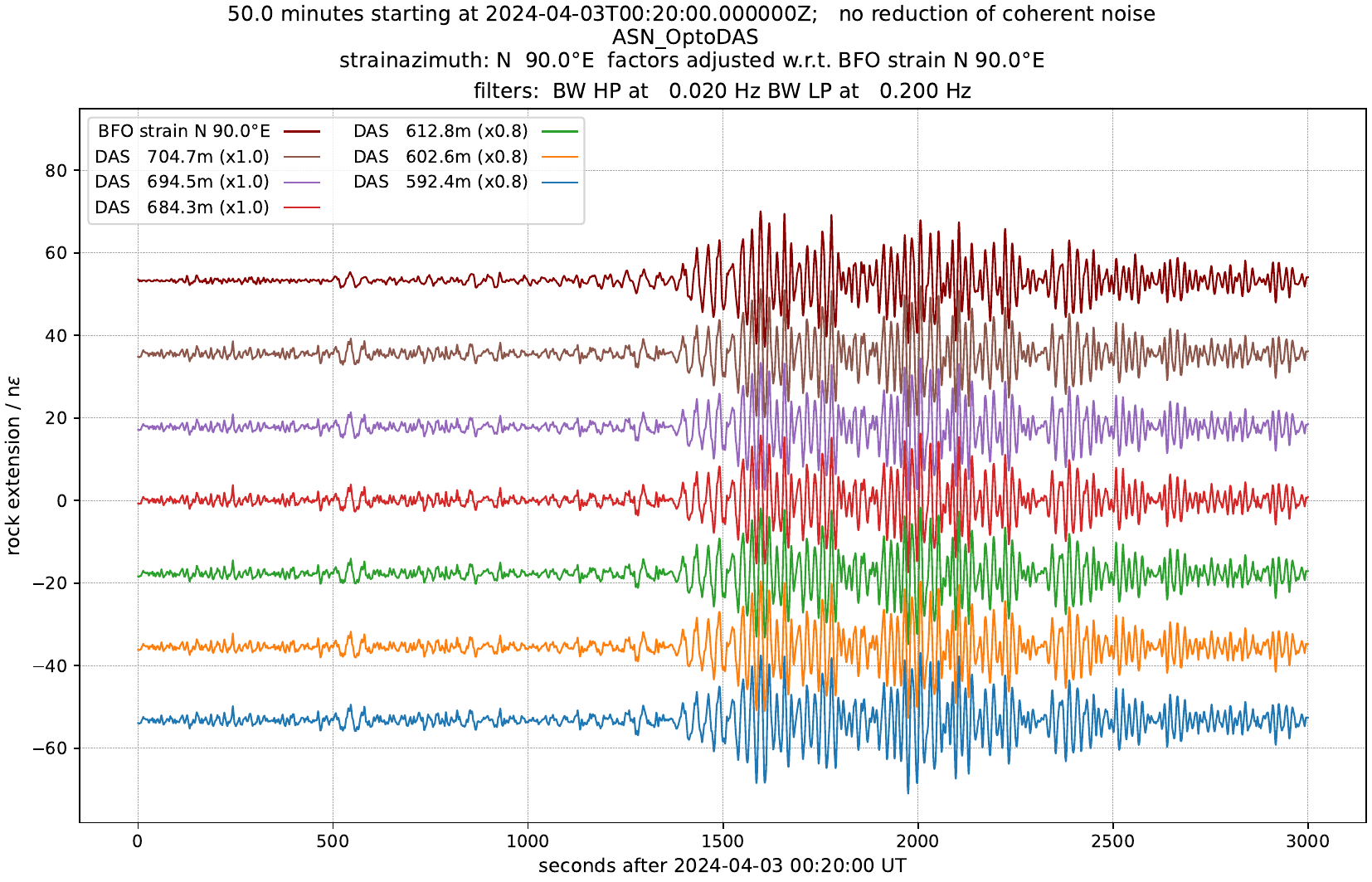}{\IUoptodas}{fig:Taiwan:OptoDAS}{}{}

\FigAlbstadtEQ{OptoDAS_ASN_OptoDAS}{\IUoptodas}{fig:Albstadt:OptoDAS}{}{\nrcomment}
\FigItalyEQ{OptoDAS_ASN_OptoDAS}{\IUoptodas}{fig:Italy:OptoDAS}{}{\nrcomment}
\FigPreSwapMMrms{OptoDAS}{ASN_OptoDAS}{\IUoptodas}{fig:MM:PreSwap:OptoDAS}{ASN
\IUoptodas}
\FigPostSwapMMrms{OptoDAS}{ASN_OptoDAS}{\IUoptodas}{fig:MM:PostSwap:OptoDAS}{ASN
\IUoptodas}

\clearpage
\subsection{Terra15 \IUtreble}
\begin{table}[h]
  \caption{Parameters used by the Terra15 \IUtreble\ interrogator when
  processing the data in the unit.
  The values are given with the number of digits as specified in the data file
  headers.}
  \label{tab:treble:parameters}
  \begin{center}
    \begin{tabular}{rl}
      \toprule
      Manufacturer: & Terra15 \\
      Model: & Treble+\\
      Firmware Version: & 6.4.0rc9  \\
      Optical wavelength: & 1550.12\,nm \\ %
      Pulse width: & equals gauge length \\
      Pulse rate: & 25499.796\,Hz\\
      Refractive index of the fiber: & 1.4682 \\ %
      Opto-elastic correction factor: & 1\\ %
      Gauge length (GL): & 49.82247633292467\,m\\
      Spatial sampling interval: & 9.80114288516551\,m\\
      Temporal sampling interval: & 0.000510432\,s\\
      Primarily recorded quantity: & velocity to be converted to strain-rate\\
      \bottomrule
    \end{tabular}
  \end{center}
\end{table}

\FigTaiwanEQ{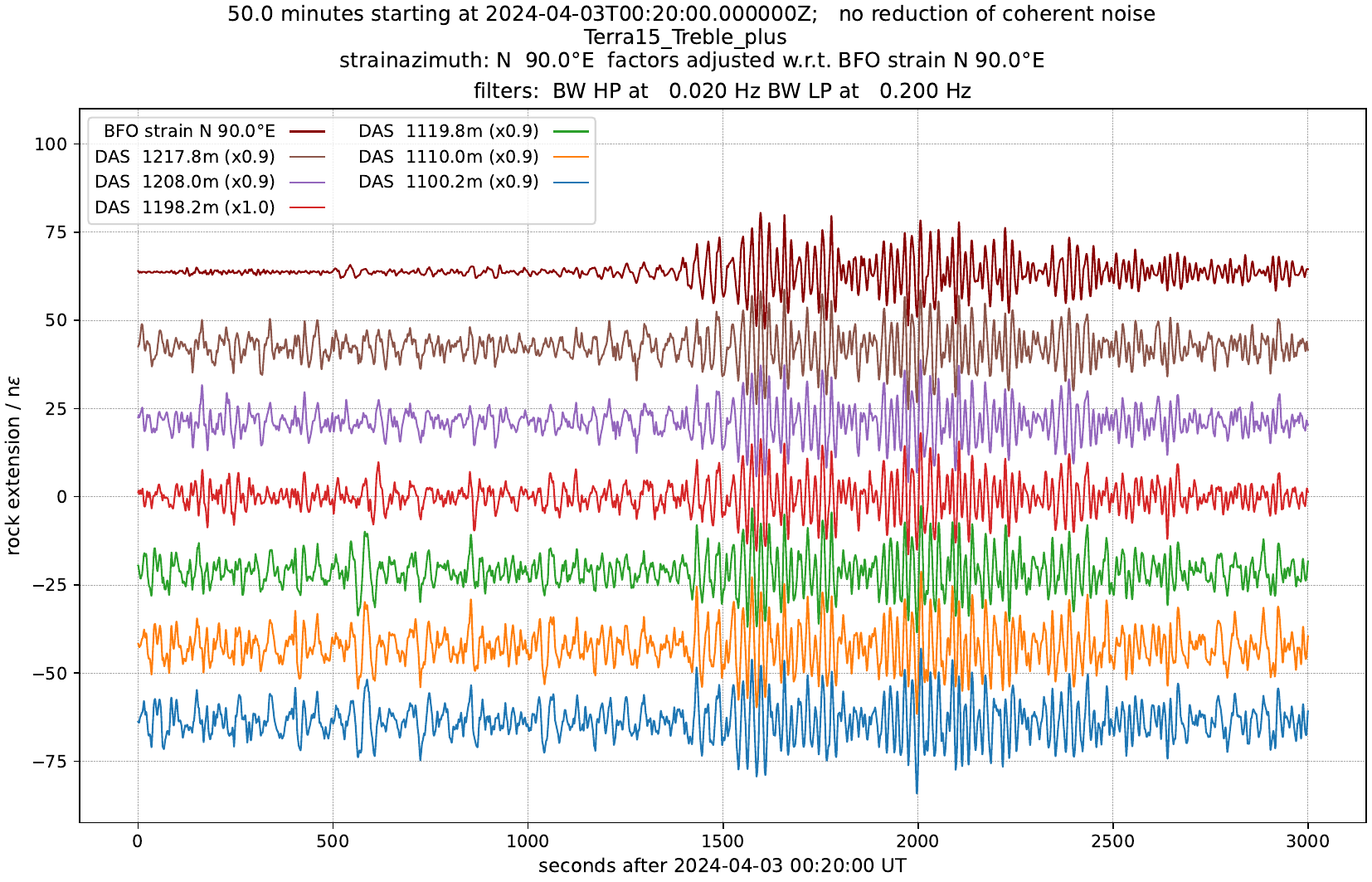}{\IUtreble}{fig:Taiwan:Treble}{}{}

\FigAlbstadtEQ{Treble_Terra15_Treble_plus}{\IUtreble}{fig:Albstadt:Treble}{_nonr}{\nonrcomment}
\FigItalyEQ{Treble_Terra15_Treble_plus}{\IUtreble}{fig:Italy:Treble}{_nonr}{\nonrcomment}
\FigPreSwapMMrms{Treble}{Terra15_Treble_plus}{\IUtreble}{fig:MM:PreSwap:Treble}{Terra15
\IUtreble}
\FigPostSwapMMrms{Treble}{Terra15_Treble_plus}{\IUtreble}{fig:MM:PostSwap:Treble}{Terra15
\IUtreble}
%

\end{document}
%
